\documentclass[a4paper,11pt]{article}
\usepackage{pos}

\title{ICRC 2023 Gamma-ray Rapporteur Talk: a quick walk through the gamma-ray universe}
 \ShortTitle{ICRC 2023 Gamma Rapporteur}

\author*[a]{Rub\'en L\'opez-Coto}

\affiliation[a]{Instituto de Astrof\'isica de Andaluc\'ia,\\
 CSIC, Granada, 18008, Spain}

\emailAdd{rlopezcoto@iaa.es}

\abstract{This document attempts to summarize the Gamma ray section of the 38th International Cosmic Ray Conference held in Nagoya. There were 387 contributions submitted to this section distributed in 22 parallel oral and three poster sessions, plus four related highlight or review talks. The information included in this contribution is a description of what was reported at the conference, that represent the state of the art of the field.}

\ConferenceLogo{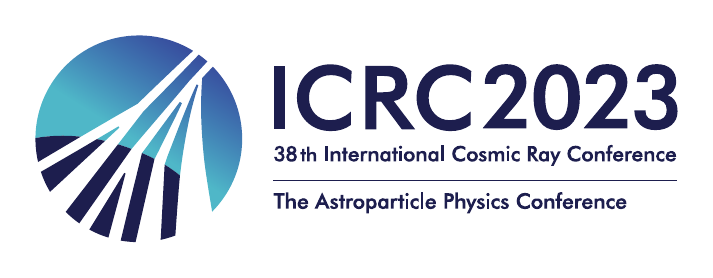}

\FullConference{%
38th International Cosmic Ray Conference (ICRC2023)\\
  26 July - 3 August, 2023\\
  Nagoya, Japan}

\begin{document}
\maketitle

\section{Instrumentation}

In the last couple of years there have been several instruments that kept producing scientific results and new ones that recently started operation and are opening new windows in the electromagnetic (EM) spectrum with unprecedented sensitivity. There are also projected instruments that have been approved or are proposed to either cover a given energy range with better sensitivity or to improve the capabilities of the detections with respect to previous instruments. This section covers all these instruments and it is divided into five different subsections with increasing energy. The techniques used to detect gamma rays in these different energy ranges are also diverse, the sub-GeV energy range being covered by balloons and satellites, the GeV to TeV accessible to Imaging Atmospheric Cherenkov Telescopes and finally the TeV to PeV dominated by Particle Detectors.

\subsection{The keV energy range} 
This energy range (also sub-MeV) is where phenomena like the Cosmic X-ray Background (CXB) or Gamma Ray Bursts (GRBs) are monitored. There were presentation of balloons like CXBe, the flexible X-ray detector focused on the CXB studies \citep{Produit:2023rup}, POLAR-2, the next generation of GRB polarization detector that will be located in the Chinese Space Station (CSS) after the successful results of POLAR \citep{Produit:2023dei}, the MoonBEAM cislunar SmallSat design  \citep{Fletcher:2023zxp} or XL-Calibur, a proposal for a balloon-borne X-ray polarimeter \citep{Uchida:2023qhk}.

\subsection{The MeV energy range}
\begin{figure}
   \centering
   \includegraphics[width=\hsize]{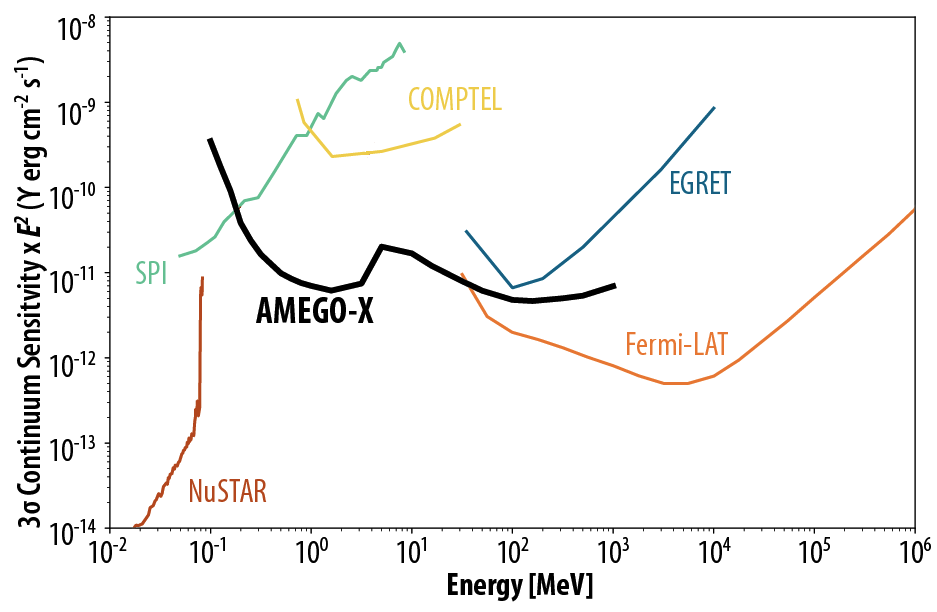}   
      
      \caption{MeV gap and instruments proposed for the future AMEGO-X \citep{Karwin:2023xbg}.
              }
         \label{fig:mev_sensitivity}
   \end{figure}

In this energy range, phenomena like positron annihilation, nuclear lines and polarization are studied. The MeV gap is the region of the EM spectrum that has not been covered by any instrument with an improved sensitivity since the launch of COMPTEL several decades ago (see Fig. \ref{fig:mev_sensitivity}). To cover this gap, there are several proposals like the the COSI Compton small explorer satellite to be launched in 2027 \citep{Tomsick:2023aue}. 
There are also balloon prototypes of more ambitious projects like ComPair \citep{Valverde:2023cyb}, that will be carrying some of the load proposed for AMEGO-X, the medium satellite explorer improving about two orders of magnitude the sensitivity in the MeV energy range \citep{Karwin:2023xbg} as it can be seen in Fig. \ref{fig:mev_sensitivity}.

This energy range will not only be covered by instruments focused on improving the sensitivity, but also on proposals to improve the angular reconstruction of the received photons, like GECCO, the Compton satellite with Coded Aperture Mask \citep{Moiseev:2023zkv}. Other instruments proposed in this energy range are balloons like SMILE-2  \citep{Ikeda:2023efo}, GRAMS \citep{Aramaki:2023ubm}, miniSGD \citep{Okuma:2023fnk}, HEPD-02, a payload of CSES \citep{Lega:2023iuj} and the XRPix detector \citep{Hashizume:2023ijl}.

\subsection{The GeV energy range}
This energy range was traditionally dominated by instruments that had been studying point-like sources as well as large scale emission for several years. We had reports from the CALET satellite at the ISS \citep{CALET:2023vld} and DAMPE \citep{DAMPE:2023amf} that have both been working for around 8 years. We also had studies of performance of polarimetry with  {\it Fermi}-LAT \citep{Laviron:2023rhx} and the latest results from the GRAINE balloon whose last flight took place in 2023 \citep{Takahashi:2023hgs}. Future proposals of satellites like HERD for the CSS \citep{HERD:2023zjq}, VLAST \citep{Zhang:2023zjq} or ADAPT, a balloon with a final goal of a super-Fermi satellite \citep{Chen:2023nij}.

\subsection{The TeV energy range}
\begin{figure}
   \centering
   \includegraphics[width=0.8\hsize]{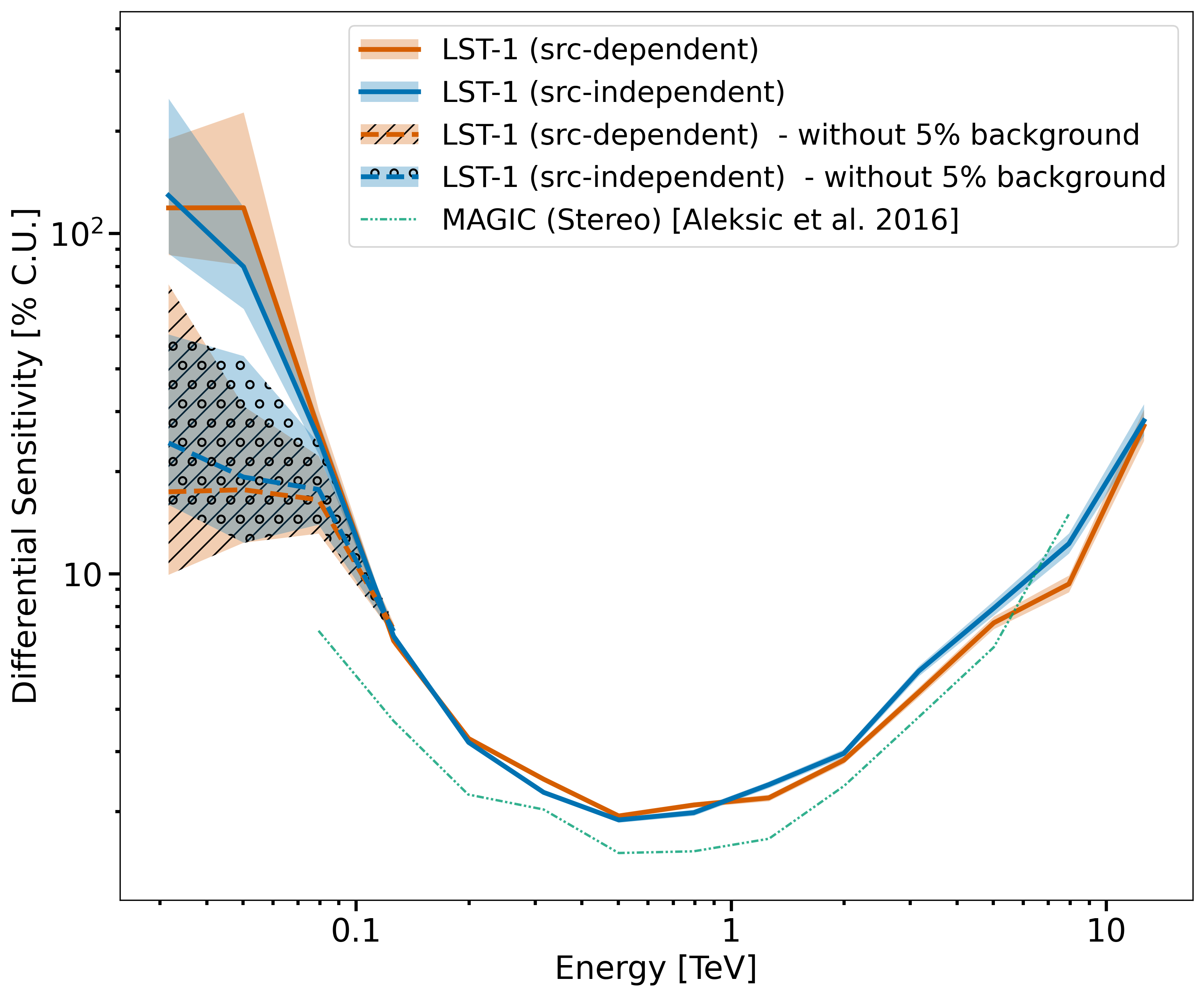}   
      \caption{Differential sensitivity of the Large-Sized Telescope of CTAO \citep{CTALSTProject:2023vhk}.
              }
         \label{fig:lst_sensitivity}
   \end{figure}

This energy range has usually been covered by imaging atmospheric Cherenkov telescopes (IACTs), instruments with excellent resolution that allows to study point sources, morphology and achieve a good spectral precision, systems like MAGIC \citep{MAGIC:2023liw} that have been operating during 20 years, presented their updates. We also had updates from instruments in construction like the ASTRI Mini-Array project \citep{Giuliani:2023smc}.
The next generation of instruments working in this energy range, that will improve the sensitivity and resolution is the CTAO Observatory. CTAO will be composed of telescopes of different sizes, the Large-Sized Telescope will be the largest of the array and the prototype is already ready and is showing its first results \citep{CTALSTProject:2023vhk}, confirming an excellent performance \citep{CTALSTProject:2023ezj} (see Fig.~\ref{fig:lst_sensitivity}). The other telescopes of the array like the Medium-Sized Telescope \citep{Bradascio:2023lmr}, Schwarzschild-Couder Telescope \citep{Kieda:2023upg} and Small-Sized Telescope \citep{CTAConsortium:2023ngo} also showed their updates, as well as the extension of the observatory with the CTA+ project \citep{Antonelli:2023uig}.

New developments in this energy range like the HADAR refracting atmospheric Cherenkov telescope \citep{Qian:2023xzx}, the MAGIC + LST-1 joint usage \citep{DiPierro:2023olc} and a proposal to connect them with a hardware trigger with the HaST system \citep{CTALSTProject:2023mpu}, the study of divergent pointing to increase IACTs field of view \citep{CTAConsortium:2023ybn}, LACT, an array of imaging atmospheric Cherenkov Telescopes for LHAASO \citep{Zhang:2023bsi} or the development of Silicon Photomultiplier cameras for LSTs \citep{Heller:2023qbh}. Finally, the usage of VERITAS \citep{Kieda:2023uig} and MAGIC \citep{MAGIC:2023ike} to perform stellar intensity interferometry measurements with very competitive performance was also shown at the conference.

\subsection{The PeV energy range}
This energy range, even though it was within reach since several years, it was only accessible to study several sources since less than a decade. It is worth highlighting the role that HAWC \citep{HAWC:2023qbx}, working for more than eight years and LHAASO \citep{Gao:2023jgz} have done extending the energy coverage. One cannot forget the role that other instruments working in the same energy range have been performing since several years ago like Tibet AS-$\gamma$ \citep{TibetASgamma:2023nks}, GRAPES-3 \citep{Pattanaik:2023mva} and the TAIGA-IACT telescopes for Multi-Messenger observations \citep{TAIGA:2023xtz}. It is worth mentioning that all the above instruments are working in the Northern hemisphere, and therefore a similar coverage in the Southern Hemisphere comes as a must, thus, the future is also very promising for instruments covering this energy range. In the Southern Hemisphere, there are arrays planned or being constructed with detectors based on scintillators like ALPACA \citep{SubietaVasquez:2023kci} or its upgrade Mega-ALPACA \citep{Sako:2023kyj}. The Southern Wide Field Gamma-ray Observatory (SWGO) \citep{Conceicao:2023tfb}, whose technology and location are still under study, has ambitious sensitivity and resolution goals to be able to investigate multi-TeV emitting sources. There are also proposals like PANOSETI, the Pulsed optical signal detector that can also be used for PeV Gamma-ray Astronomy \citep{Korzoun:2023jgb}.

\section{Galactic science}

One of the most important unanswered questions in astroparticle physics as of now is that of the sources accelerating hadronic CRs up to PeV energies, the so-called PeVatrons. It is widely agreed that these sources need to have Galactic origin, but there are a zoo of different sources modeled to accelerate particles up to PeV \citep{Fang:2023ugh}, we will discuss some of the most important candidates as well as sources of leptonic CRs known to have the capabilities to produce PeV particles.

\subsection{Supernova Remnants}

\begin{figure}
   \centering
   \includegraphics[width=\hsize]{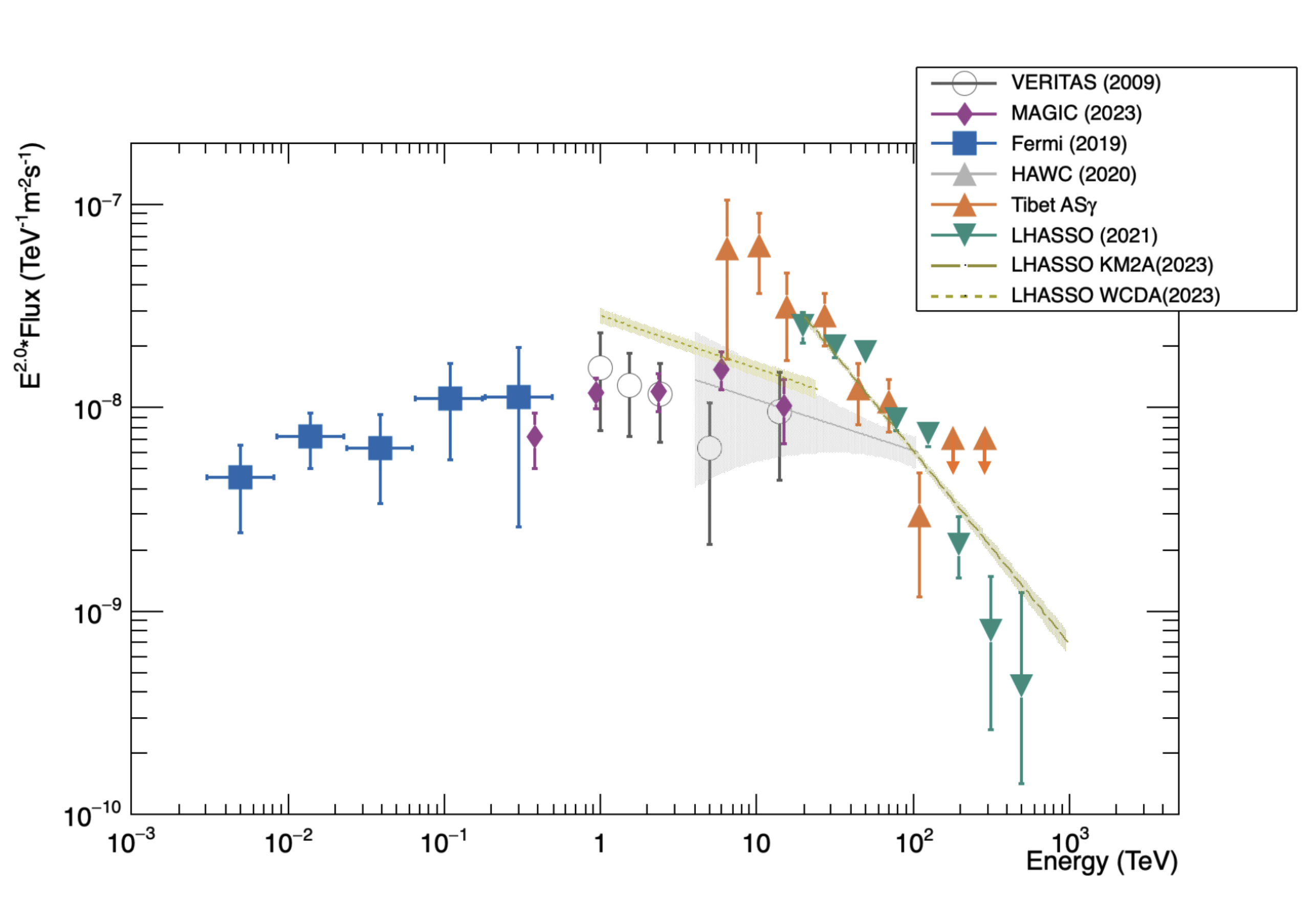}   
      \caption{Boomerang spectrum including data from different experiments \citep{Park:2023bsm}.
              }
         \label{fig:boomerang_spectrum}
   \end{figure}

The usual suspects are Supernova Remnants (SNRs) due to the fact that they provide in the Galaxy enough power to account for all galactic cosmic rays with only a fraction of their total energy budget. Some already known sources for which we had some updates are W51C, measured by LHAASO up to 300 TeV \citep{Chen:2023tpu} for which still hadronic models (with an energy cut-off at 400 TeV) are favored to explain the emission. Updates on the measurement of another claimed PeVatron SNR G106.3+2.7 (Boomerang) were shown by HAWC \citep{HAWC:2023sfg}, MAGIC \citep{MAGIC:2023kkn} and also VERITAS \citep{Park:2023bsm}, showing a spectrum that extends up to 500 TeV (see Fig. \ref{fig:boomerang_spectrum}). The VHE observations of $\gamma$ Cygni \citep{Feng:2023daw, MAGIC:2023kkn} favor a hadronic origin for the emission although also reaching lower than PeV energies. New sources like LHAASO J2002+3238 is spatially associated with SNR G69.7+1.0 \citep{Hou:2023rte} but the origin of the VHE $\gamma$-ray emission is unclear whether it is leptonic or hadronic. SNR G150.3+4.5 observed by LHAASO \citep{Zeng:2023uvu} is a SNR with a pulsar at the center for which both leptonic and hadronic scenarios work. Unfortunately the TeV Morphology of the source is still under study and whenever settled, it may help distinguishing scenarios.

\begin{figure}
   \centering
   \includegraphics[width=\hsize]{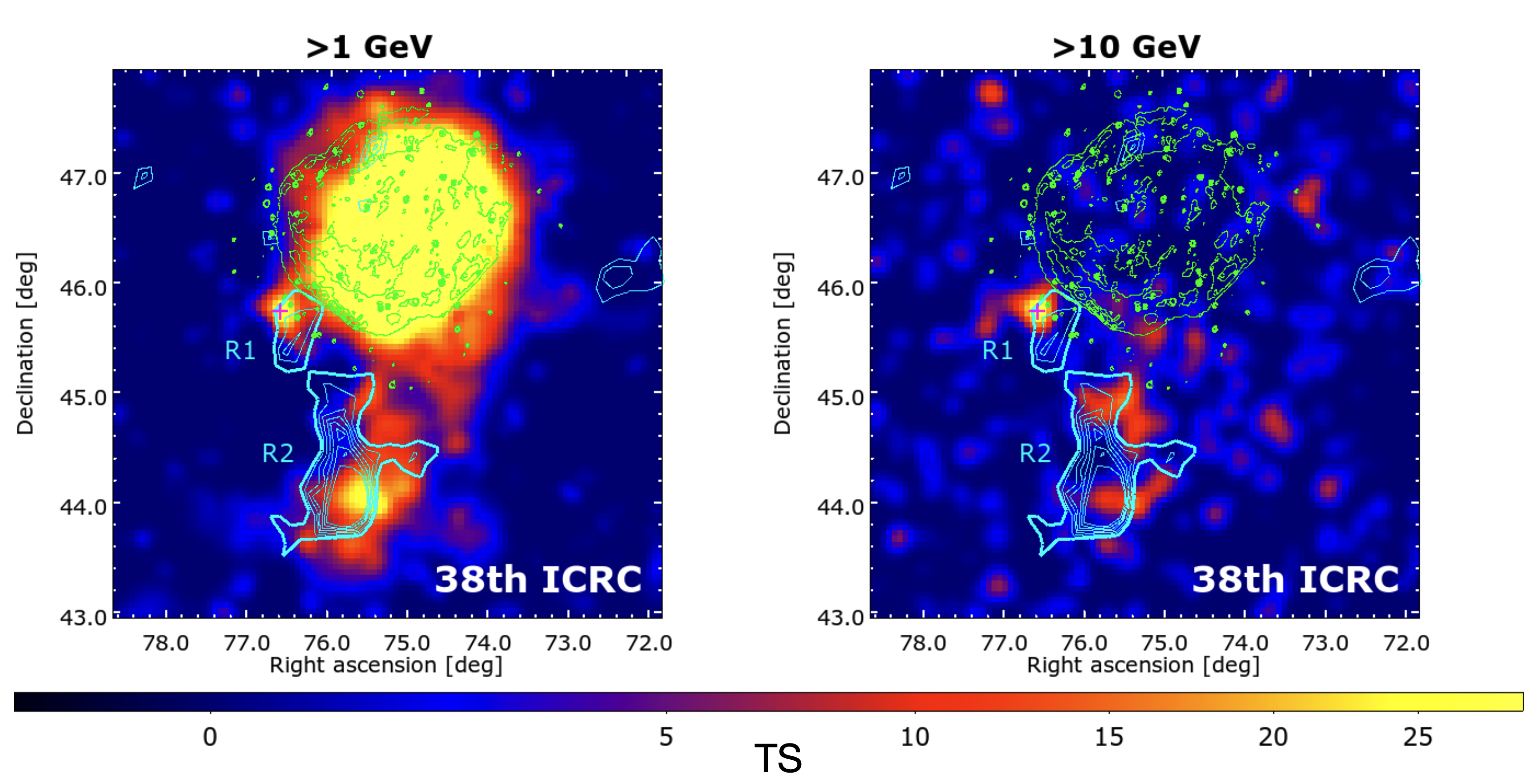}   
      \caption{HB9 Skymap above 1 GeV (left panel) and 10 GeV (right panel). Radio continuum 1420 MHz are shown in green and $^{12}$CO (J=1-0) in cyan \citep{Oka:2023vaw}.}
         \label{fig:hb9}
   \end{figure}

Other SNRs accelerating protons to lower energies are, for example, the HB9 delayed emission due to protons illuminating a nearby molecular cloud by {\it Fermi}-LAT \citep{Oka:2023vaw} (see Fig. \ref{fig:hb9}) or HB3 \citep{Clement:2023sbx} and Puppis A \citep{Giuffrida:2023udf, Aruga:2023udf} detections by the same satellite. Finally, modeling of these sources is very important in order to reproduce their morphology and spectra as it wass shown for W28 \citep{Einecke:2023njn} and RXJ1713.7-3946 \citep{Rowell:2023ypc} 

\subsection{Stellar Clusters}
Since PeV proton acceleration is difficult to be achieved for SNRs, people started to look for other source types being able to reach PeV energies and with enough energy budget to be able to explain the energetics observed for the local CR spectrum. Stellar Clusters, with their strong winds due to the abundance of OB stars, were other obvious candidates. Since several years there has been an increase interest to investigate the origin of their VHE $\gamma$-ray emission through spectral and morphological studies. There are several models manage to get particles up to PeV energies \citep{Celli:2023nqg, Harer:2023xdq} and prospects to search for their signatures \citep{Mitchell:2023sxq} not only in gamma rays, but also in neutrinos. For sources like the  Cygnus Cocoon \citep{Guevel:2023bhg} it was shown that synchrotron emission limits the lepton contribution at the highest energies to be less than 25\% emission beyond PeV in LHAASO data \citep{Wu:2023ljn}. Westerlund 2, on the other hand, shows several components aligning with molecular structures \citep{Holch:2023jhb}. For W43, the $\gamma$-ray emission is likely generated by massive stars accelerating CRs and interacting with gas \citep{Wang:2023dsw}.

\subsection{The Galactic Center}
The Galactic Center is a very complex region with several sources possibly contributing to the VHE $\gamma$-ray emission in the region. PeVatron acceleration was claimed in the past \citep{2016Natur.531..476H} by H.E.S.S. and in this conference we had a revisiting of the region with the usage of LST-1 and MAGIC \citep{CTALSTProject:2023njo}. The correlation of the gamma rays with gas tracers is very complicated in the region and actually any modelling needs to include 3D components to be able to properly reproduce the $\gamma$-ray emission \citep{Dorner:2023rul}. Due to the mass quantity in the region, there is also the possibility of a non-negligible quantity of dark matter as it was searched for by VERITAS \citep{Ryan:2023yzu} or DAMPE \citep{DAMPE:2023aff}. Finally, structures like the Fermi Bubbles that have not been seen at higher energies up to now were also searched for by instruments like LHAASO \citep{Zhang:2023xxp}.

\subsection{Diffuse emission and Cosmic Rays}
LHAASO recently reported the study of diffuse emission in the inner galactic plane using both the KM2A and the WCDA detectors \citep{Li:2023dpg,Zhang:2023caw}. It was also also claimed that it was difficult to explain this emission using only standard CR propagation. Several works were presented in this respect in which they claimed that depending on the parameters of the propagation, the emission from the CR sea or with the contribution from sources may be the answer to explain this new data \citep{DelaTorreLuque:2023usg, Kaci:2023iie, Giacinti:2023ljr, Stall:2023hns, Abounnasr:2023ufg, Zhang:2023zww, Menchiari:2023qim}.

\subsection{Pulsar Wind Nebulae}
Even though most likely accelerating leptons, Pulsar Wind Nebulae (PWNe) like the Crab Nebula need to be accelerating electrons and positrons up to PeV energies to produce the broadband emission observed and polarized X-rays as seen by IXPE \citep{Mizuno:2023qyf}. This acceleration can be explained by models like the one presented in \citep{Giacinti:2023tde}, but there is also the possibility of a hadronic contribution to the VHE gamma-ray spectra \citep{Spencer:2023oqh}. Moreover, the Crab Nebula also shows flares at GeV energies as seen by {\it Fermi}-LAT and AGILE, and a very extensive study whose conclusion is that GeV flares mary not be driven by a single mechanism was also shown \citep{Tsirou:2023fge}.

Another interesting PWN is HESS J1825-137 viewed by LHAASO \citep{You:2023jtj} and HAWC \citep{HAWC:2023fdd}. They measured spectra ranging from a few TeV up to above 200 TeV showing an energy-dependent extended morphology that points towards the presence of electrons injected by the central source and being cooled down. For HAWC J2031+415, the HAWC morphological studies do not show any energy-dependence \citep{Herzog:2023wmj}, but more data are needed to support the PWN scenario. Additionally, the magnetar wind nebula around Swift J1834.9-0846 \citep{Li:2023ryl} claimed to be the first magnetar wind nebula powered by the internal magnetar energy of the central source.

\subsection{Halos}
\begin{figure}
   \centering
   \includegraphics[width=0.8\hsize]{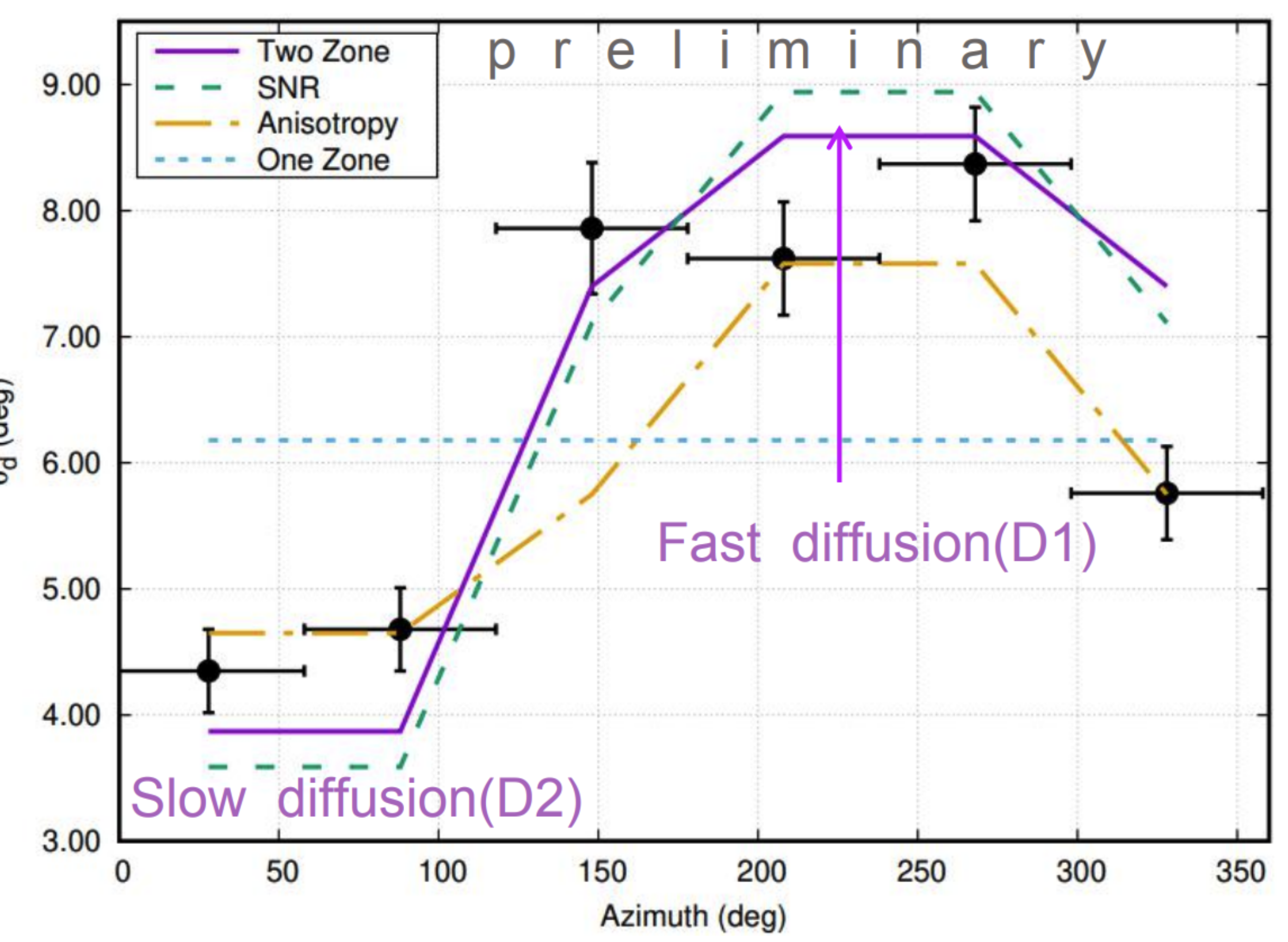}   
      \caption{Geminga VHE $\gamma$-ray emission observed by LHAASO-KM2A \citep{Chen:2023ffo}.
              }
         \label{fig:geminga_lhaaso}
   \end{figure}

Halos are regions of slow propagation glowing in gamma rays powered by pulsars \citep{Bi:2023nxr} in general dominated by diffusion \citep{Liu:2023cjn} (although there are models that claim that the mechanism is in dispute \citep{Wu:2023lgn}). Geminga and Monogem were the first confirmed sources of this type \citep{2017Sci...358..911A} and there were updates by HAWC \citep{HAWC:2023bfh} and H.E.S.S. \citep{Mitchell:2023rrd} presented at the conference. Most importantly, there was a new measurement of the emission around Geminga, this time using LHAASO-KM2A \citep{Chen:2023ffo} in which there are claims of asymmetric diffusion as it is shown in Fig. \ref{fig:geminga_lhaaso}. This could be related to the alignment of the magnetic field in the region or that that the propagation is happening inside/outside of the SNR in different regions.

\begin{figure}
   \centering
   \includegraphics[width=0.6\hsize]{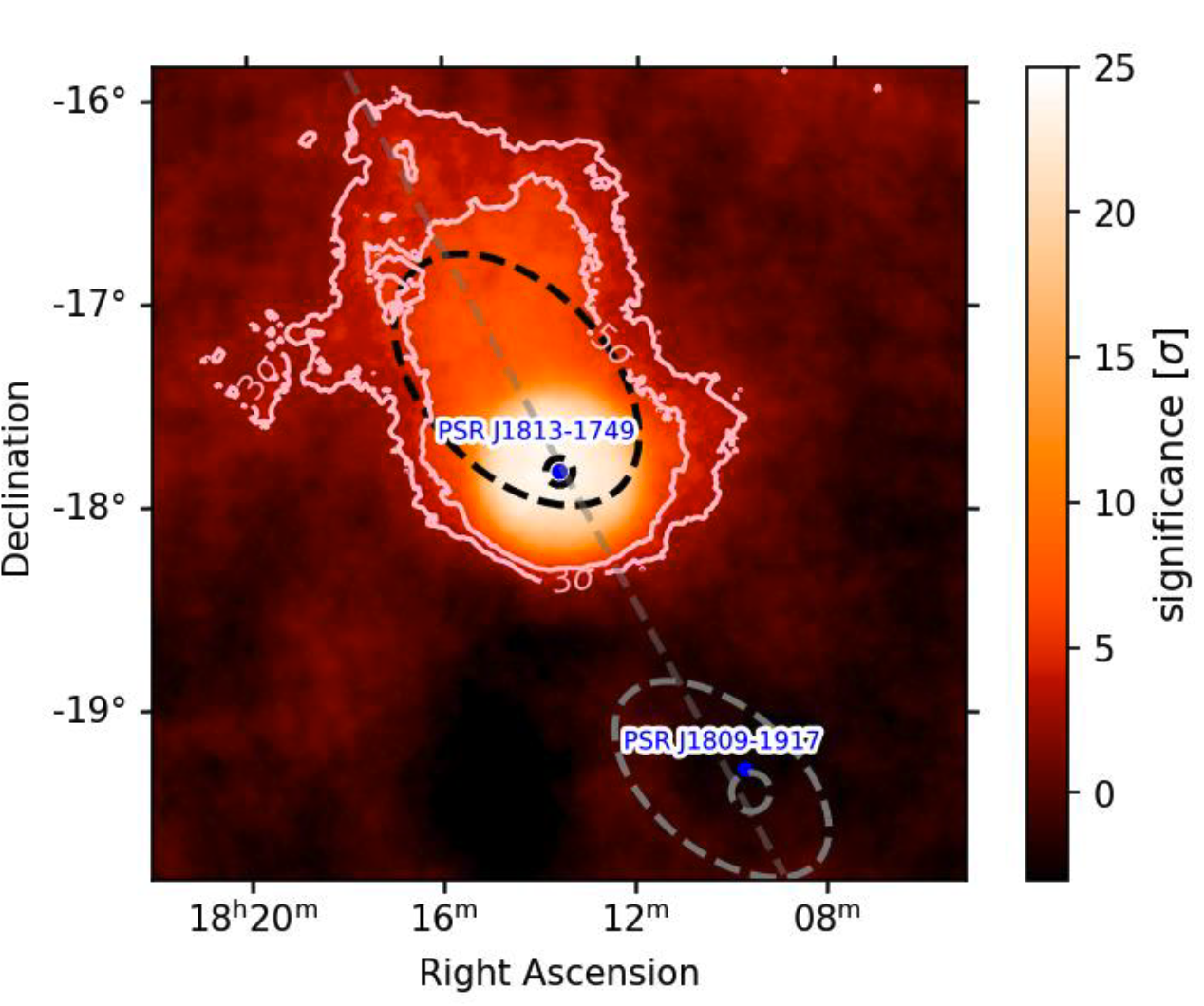}   
      \caption{PSR J1813-1749 extended emission above 0.4 TeV seen by H.E.S.S.. 
              }
         \label{fig:PSR_J1813-1749}
   \end{figure}

The extended emission surrounding PSR J1813-1749 (see Fig. \ref{fig:PSR_J1813-1749}) , despite the young age of the pulsar at the center, shows characteristics of a halo \citep{HESS:2023owq}. There are also other studies like the LHAASO measurement of the PWN tail of PSR J1740+1000 \citep{Xu:2023jtl} that might point towards a halo origin, but pulsar offset and small extension may make the scenario not suitable. HAWC J0359+5414 was put forward as a halo candidate \citep{Coutino:2023fsf} with two powerful pulsars in the region may power it. Additionally, HAWC searches for halos around pulsars \citep{Coutino:2023gsf}, millisecond pulsars \citep{HAWC:2023bfj} and M31 \citep{HAWC:2023fsz} were also shown. Finally, we also had prospects for CTAO \citep{Eckner:2023jiz} and eROSITA \citep{Li:2023fkj}.

\subsection{Pulsars}
\begin{figure}
   \centering
   \includegraphics[width=\hsize]{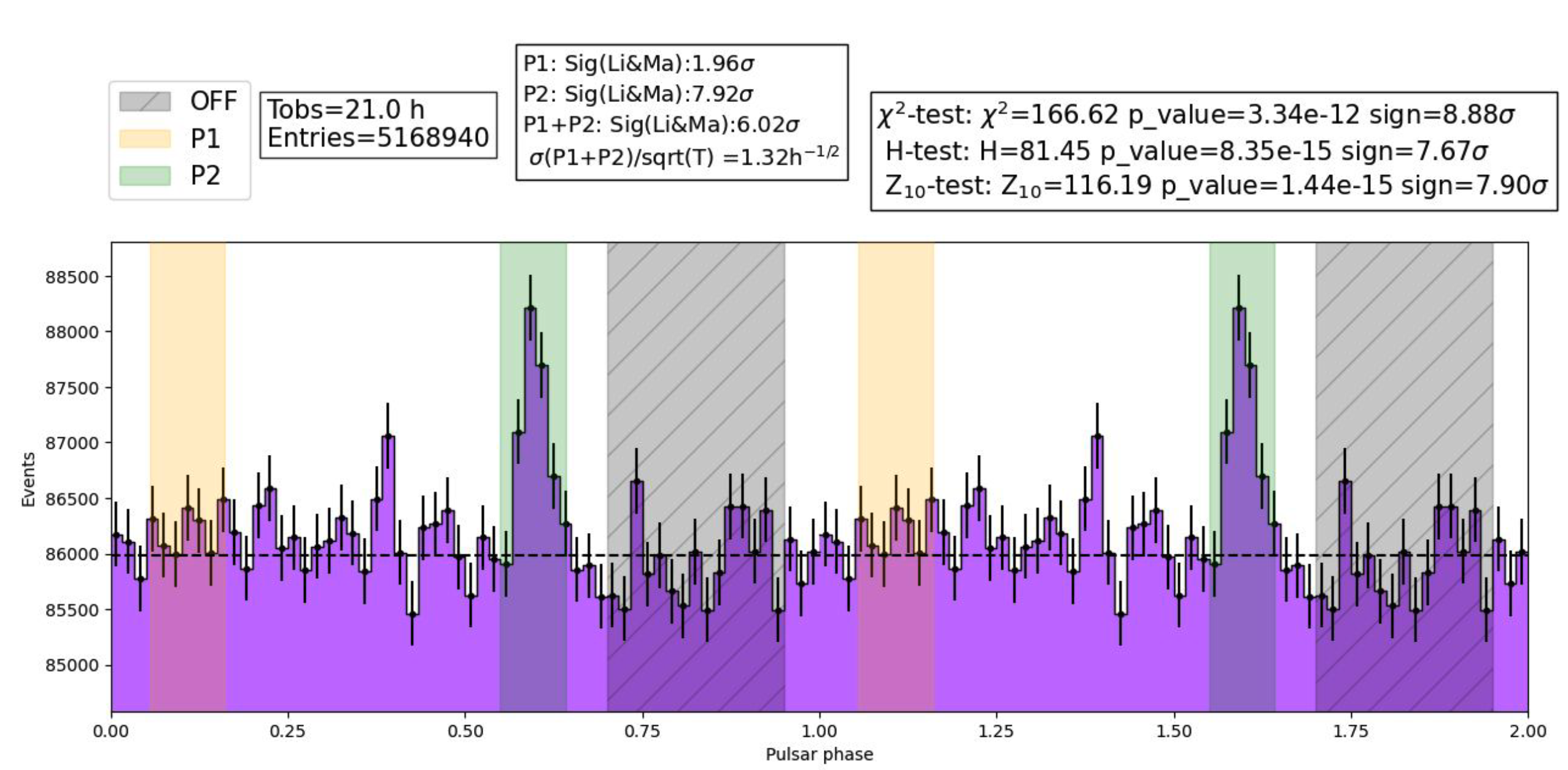}   
      \caption{Geminga pulsar phaseogram.
              }
         \label{fig:geminga_pulsar}
   \end{figure}

There are three pulsars that have been detected in the VHE $\gamma$-ray band. Two of those (Geminga and Crab) have already been detected with the LST-1 \citep{CTALSTProject:2023ekj}. Both peaks observed in the Crab and Geminga detected with a high significance as it can be seen in the phaseogram shown in Fig. \ref{fig:geminga_pulsar}. LHAASO observed gamma rays from the location of the millisecond pulsar J0218+4232 \citep{Li:2023uvu}, but even though spatially coincident, it is difficult to associate it with the millisecond pulsar and MAGIC did not find any emission coming from the region \citep{MAGIC:2023kkn}. Finally, a study of the X-ray and $\gamma$-ray emission of the variable $\gamma$-ray pulsar PSR J2021+4026 \citep{Razzano:2023gnk} observed in different wavelengths shows a gamma-to-X shift of 0.21 in phase related to mode change.

\subsection{Binaries}

\begin{figure}
   \centering
   \includegraphics[width=0.6\hsize]{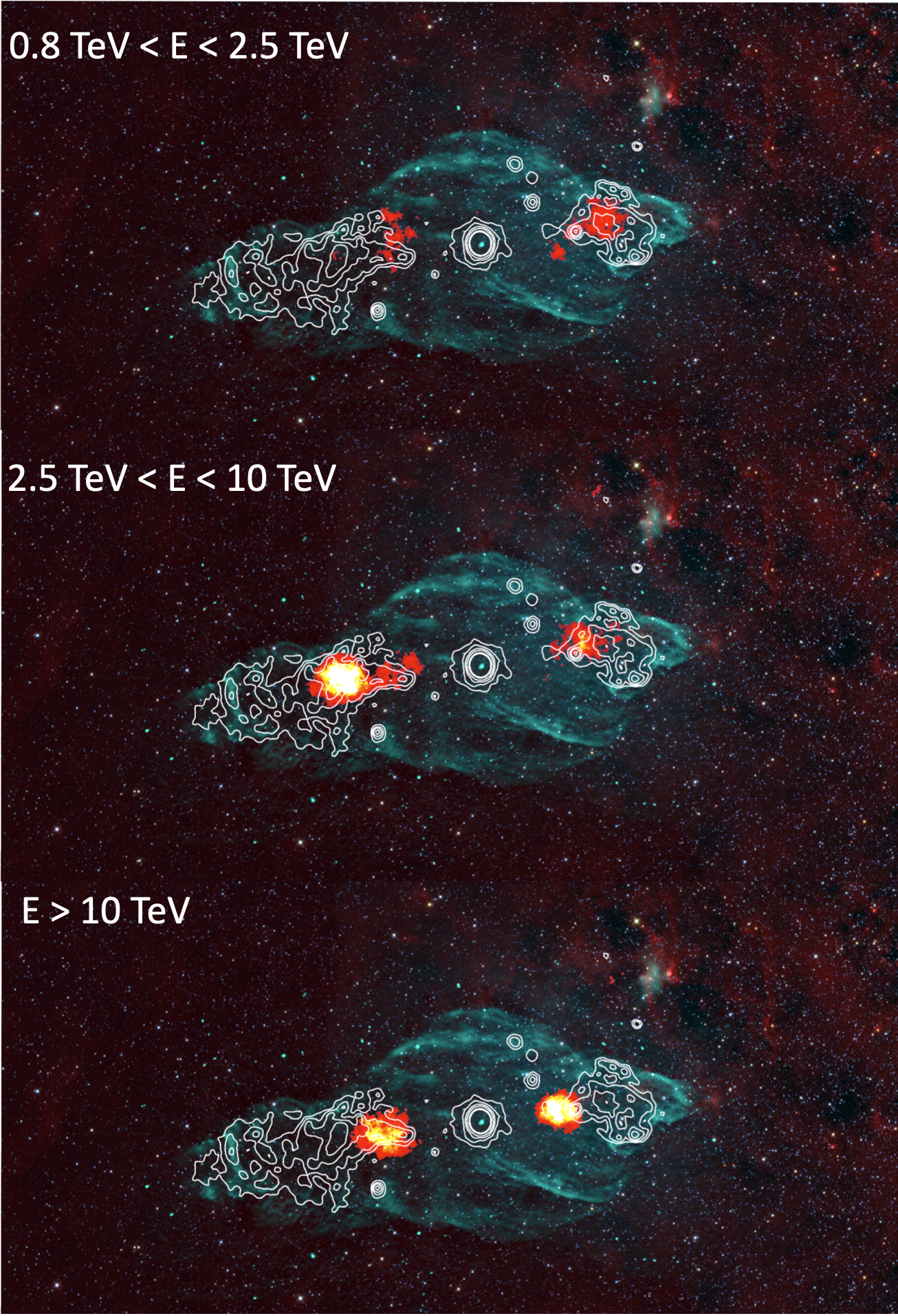}   
      \caption{SS 433 Energy dependent morphology \cite{Olivera:2023ljn}.
              }
         \label{fig:3panel_ss433}
   \end{figure}

The microquasar SS 433 was first detected by HAWC and deeply studied by H.E.S.S., that favors a leptonic origin of the emission \cite{Olivera:2023ljn}. The electrons propagate from the central source and cool down faster for higher energies as it is depicted in Fig. \ref{fig:3panel_ss433}. HAWC also updated its original results with a well-measured spectrum \cite{HAWC:2023dtw} in contrast with the original report of only one flux point. Other results on the study of binary systems were the observation of the periastron passage of PSR B1259-63 \citep{Thorpe:2023lrg}, composed by a pulsar and a O9.5Ve-type star binary system with equatorial disk that the pulsar crosses twice. It showed X-ray-to-TeV correlation more strongly after second disc crossing, but no GeV-to-TeV correlation. The $\gamma$-ray binary LMC P3 showed the peak of emission after inferior conjunction as seen by H.E.S.S. \citep{Fisher:2023bdg}. We could also see the VHE gamma-ray observations \citep{HESS:2023nna} and modeling \citep{Walter:2023hfl} of Eta Carina, the MAGIC observation of HESS J0632+057 and MAXI J1820+070 \citep{MAGIC:2023kkn}, the LS 5039 modulation and source coincident with V4641 Sgr detected by HAWC \citep{HAWC:2023acx} and the observation of the Be/X-ray binary LS V +44 17 by VERITAS during outburst \citep{Holder:2023tlg}.

\subsection{Novae}

\begin{figure}
   \centering
   \includegraphics[width=\hsize]{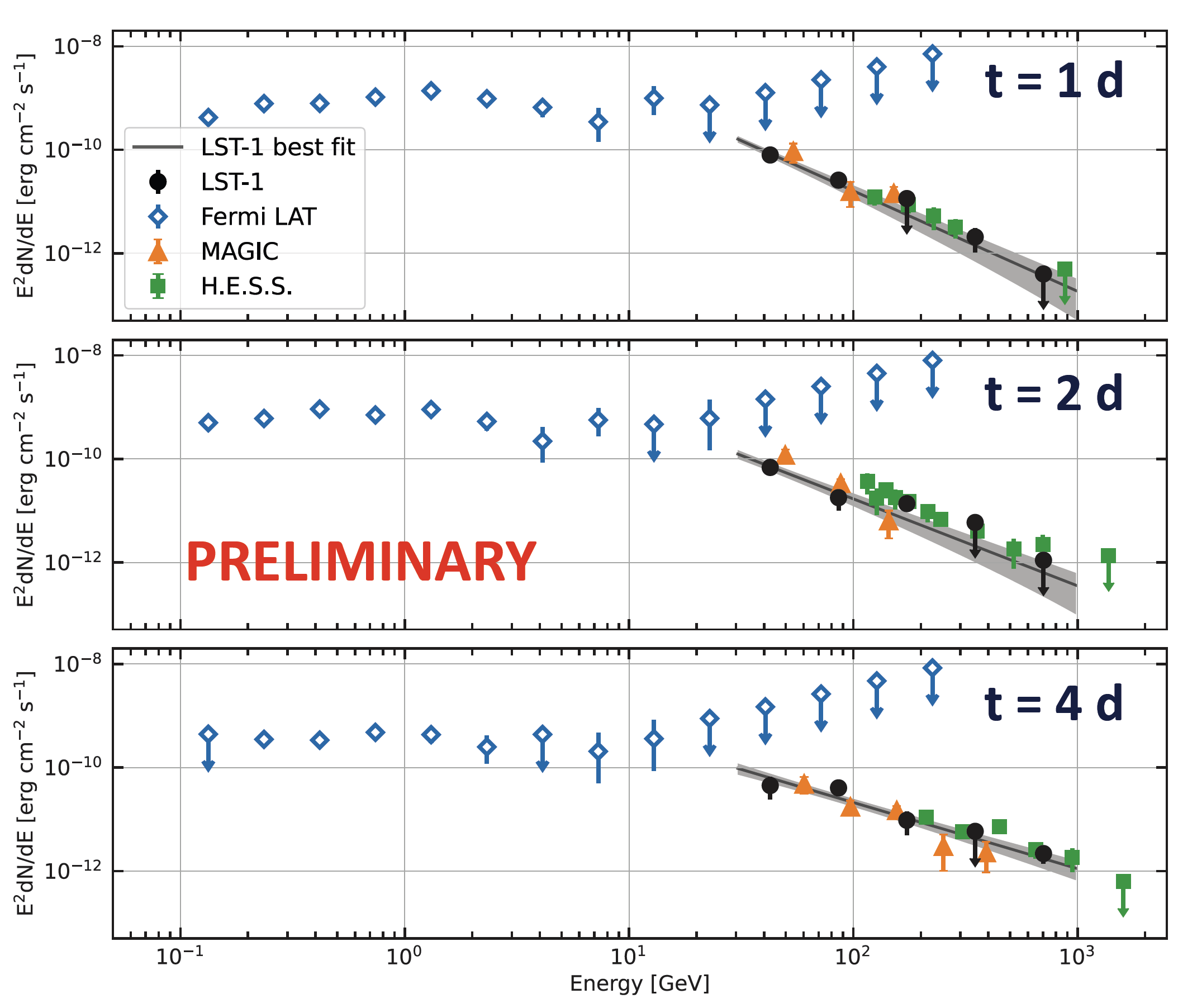}   
      \caption{Daily RS Oph spectra \citep{CTALSTProject:2023mgl}.
              }
         \label{fig:daily_rsoph_spectra}
   \end{figure}

Although novae had been established as gamma-ray emitters several years ago, they had never been detected in the VHE $\gamma$-ray band until RS Ophiuchi, a symbiotic nova that erupted August 8th 2021 was detected by H.E.S.S., MAGIC and the LST-1 for several days after the eruption (see Fig. \ref{fig:daily_rsoph_spectra} for the evolution). Proton acceleration is strongly favored \citep{MAGIC:2023wxo, CTALSTProject:2023mgl} to explain the $\gamma$-ray emission and these accelerated protons will eventually escape nova shock and contribute to the sea of CRs. Although the previous works using simple modeling assume that the same particle population produce the GeV-to-TeV emission, there are other models that disfavor the single-shock scenario to explain it \citep{Diesing:2023chh}.

\subsection{Galactic Transients}

There are a plethora of source types that might produce transient $\gamma$-ray emission \citep{Hadasch:2023ljn}. There are other candidates for PeV cosmic-ray acceleration like for example core-collapse supernovae in dense environments. Transitional millisecond pulsars whose origin of the gamma-ray emission during the intermediate accretion stage is unclear or magnetars and Fast Radio Bursts searches with VERITAS \citep{HAWC:2023gcx}, H.E.S.S. \citep{Jaitly:2023fcx}, HAWC \citep{Willox:2023lsf}, {\it Swift} and {\it Fermi}-LAT  \citep{Ashkar:2023omc}.

\subsection{Unidentified sources}

There are still sources for which the nature of their emission is still unclear and they are therefore classified as unidentified. Below we can find a list of them with a few notes that may help to identify them in the future or understand the origin of their current classification.

\begin{itemize}
\item MGRO J1908+06. Leptonic model favor to explain the VHE gamma-ray emission observed by VERITAS \citep{Kleiner:2023ehi}.
\item LHAASO J0341+525. Emission confirmed by HAWC \citep{Bangale:2023ktt}
\item LHAASO J1959+2850. Powerful pulsars in the surroundings, therefore it is likely a PWN  \citep{Yu:2023jia}.
\item LHAASO J1929+1745 region \citep{Guo:2023jia}. It shows a similar morphology than that in HAWC \citep{HAWC:2023qbx}
\item LHAASO J2108+5157. Observed by LST-1 \citep{CTALSTProject:2023bfh} and VERITAS+HAWC \citep{Kumar:2023txz}. Molecular clouds in the surroundings \citep{Toledano:2023ncz} may point towards a hadronic origin of the emission, although modeling shows a compatibility between the origin being leptonic as well.
\item HAWC counterpart of LHAASO J1849-0003 \citep{HAWC:2023dsl}. There is a high spin-down power pulsar in the region, and therefore the origin is expected to be of a PWN.
\item HESS J1809-193. Observed by HAWC \citep{HAWC:2023lst}, the lepto-hadronic scenario is favored by the modelling, but a halo could also be the origin as indicated by X-rays from Chandra \citep{Li:2023mia}.
\end{itemize}

\section{Extragalactic science}

\subsection{Starburst Galaxies}

Starburst Galaxies are factories of cosmic rays that glow in gamma rays. The origin of their emission is thought to be due to the strong winds from the stars contained in the galaxy. In this conference, we had an update on the M82 detection by VERITAS\cite{Saha:2023ent} and also the proposal of PWNe as the origin of the $\gamma$-ray emission and their non-negligible contribution to the Extragalactic Gamma-Ray Background \cite{Owen:2023sqe}


\subsection{Blazars}
Active Galactic Nuclei (AGNs) present different features depending on the angle their ultrarelativistic jets form with the line of sight from the Earth. Blazar jets point directly towards the Earth. We had several presentations about the theory and prospects, like those studying the nature of TeV $\gamma$-ray variability in blazars \citep{Baghmanyan:2023aqf, Lindfors:2023efw}, SED modeling \citep{Boula:2023upv} and second-order effects \citep{Dmytriiev:2023wkf}.

The theory for the emission of low-luminosity AGNs is advancing with studies like the spine-sheath jet model \citep{Boughelilba:2023fet} or that of the emission from their jets \citep{Tomar:2023wzn}. For the future we also had a novel method to identify blazar emission states using clustering algorithms \citep{Heckmann:2023qtg} and prospects for CTAO with the study of bright flares \citep{CTAConsortium:2023grr}, variability predictions \citep{Grolleron:2023dtw} and redshift determination \citep{Lindfors:2023ffw}.

\begin{figure}
   \centering
   \includegraphics[width=\hsize]{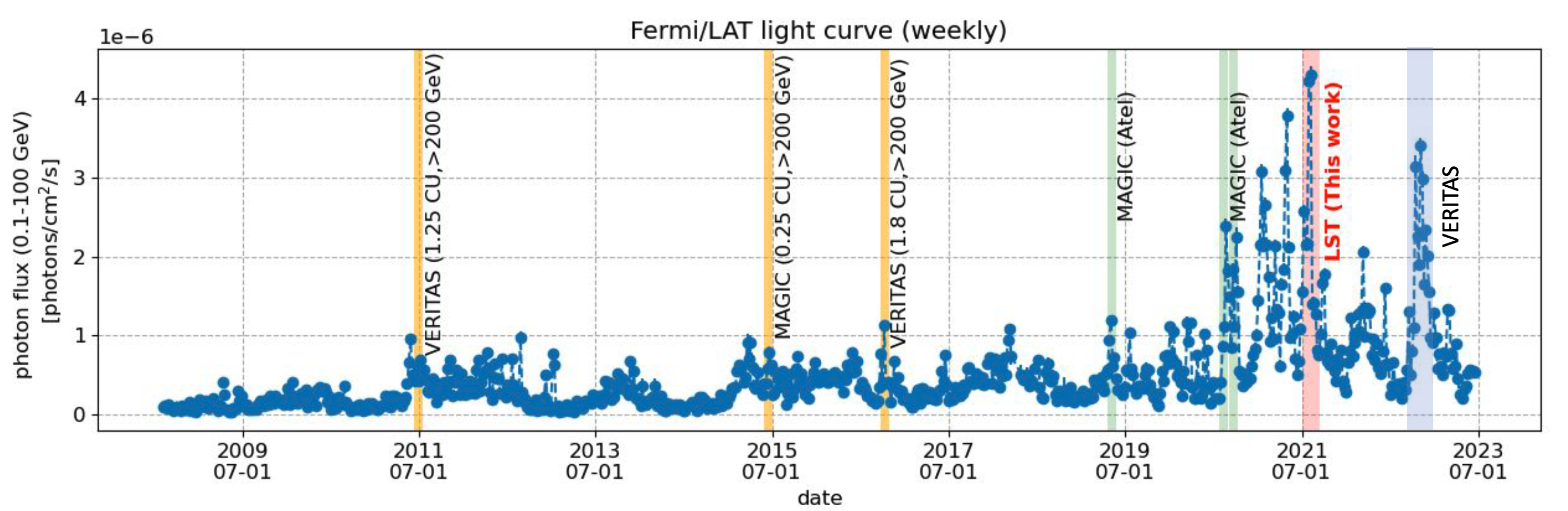}   
\caption{BL Lac {\it Fermi}-LAT long-term lightcurve evidencing the 2020, 2021 and 2022 flares together with other episodes\citep{CTALSTProject:2023aoh}.}
\label{fig:bllac_flares}
\end{figure}

BL Lac is the prototypical blazar and it shows very frequent flares (see Fig. \ref{fig:bllac_flares}) as those observed by MAGIC in 2020 \cite{Imazawa:2023ezy}, LST-1 in 2021 \cite{CTALSTProject:2023gmi} and VERITAS in 2022 \cite{Hinrichs:2023vjm}, every one of them evidencing a different aspect of the emission.

Mrk 421 and 501 are some of the most monitored blazars in the VHE $\gamma$-ray sky. We had updates on the observations by HAWC \citep{HAWC:2023szn} and LST-1 \citep{CTALSTProject:2023aoh}, added to the X-ray and TeV correlation study from Mrk 421 by VERITAS \citep{Mooney:2023ppi} that favors leptonic models and the Lorentz Invariance Violation constraints using Mrk 421 as well \citep{Damico:2023rup}. Kinks may be present in the spectra of these sources, as it hinted in the past, which may point to structured jets \citep{Becerra:2023lrb}.

\begin{figure}
   \centering
   \includegraphics[width=\hsize]{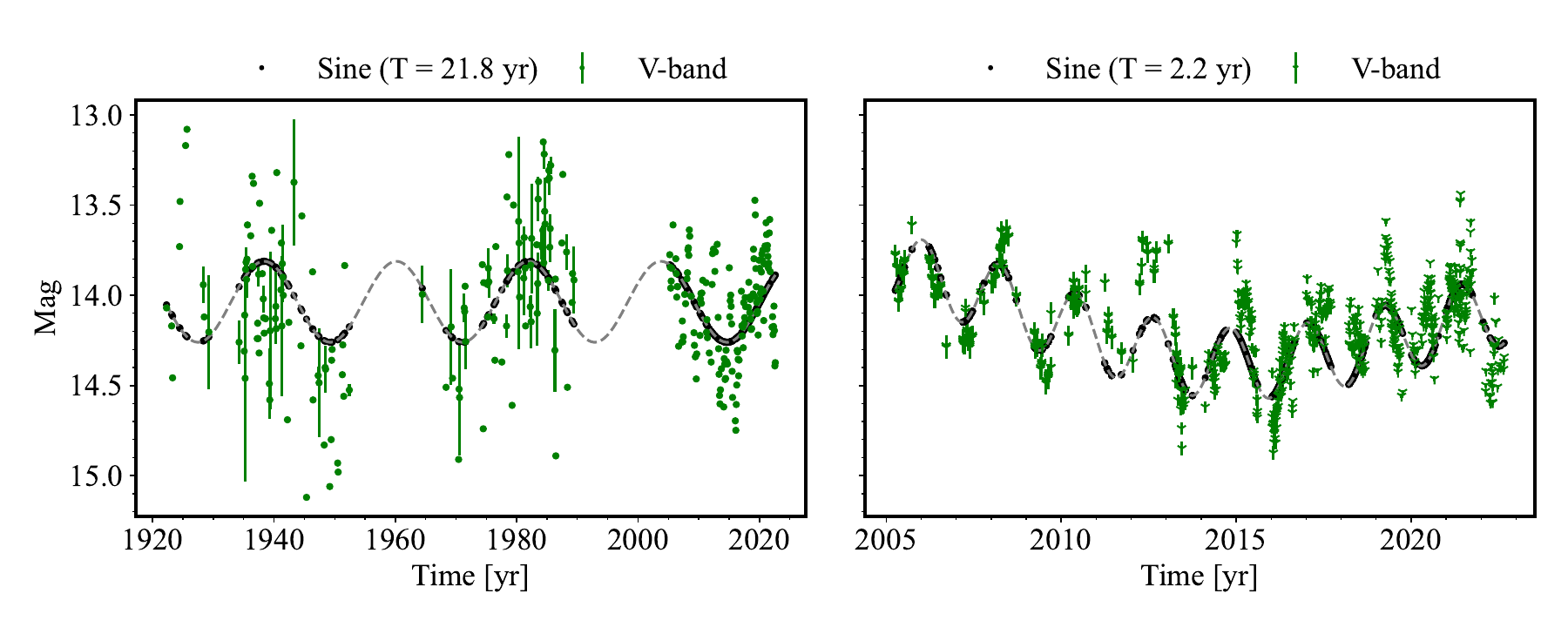}   
\caption{PG 1553+113 2.2 (right panel) and 21.8 year period (left panel) derived using optical data in the V-band \citep{Dominguez:2023enu}.}
\label{fig:PG1553+113}
\end{figure}

PG 1553+113 has been claimed to contain a binary black hole at the center \citep{Dominguez:2023enu} because of the evidence for a 2.2 yr periodicity that can be seen in {\it Fermi}-LAT data. In this contribution, the hypothesis that was put forward was that the 2.2 yr periodicity of the $\gamma$-ray data can be part of a longer trend with a long-term period of $\sim$22 yr that has a 3$\sigma$ statistical significance when considering long-term optical data in the V-band (see Fig. \ref{fig:PG1553+113}). A lump scenario was put forward as the explanation for this additional periodicity.

Finally, there are monitoring programs of several AGNs, like that shown in the VERITAS AGN Highlights \citep{Benbow:2023maj}: RBS 1366 \citep{Ribeiro:2023jgj} and S3 1227+25 \citep{Acharyya:2023ogc}, flaring Blazar ToOs with H.E.S.S. \citep{HESS:2023mnr}, HAWC detection of Mark 421, 501, M87 and 1ES 1215+303 \citep{HAWC:2023gto}, the Flat Spectrum Radio Quasars monitoring with MAGIC \citep{MAGIC:2023zrc} or the MAGIC detection of GB6 J1058+2817 and B2 1811+31 \citep{MAGIC:2023hrz}.

\subsection{Radio Galaxies}
 Radio Galaxies are AGNs whose jet subtends a larger angle than blazars with respect to the line of sight from the Earth. M87 is one of these long-known radio galaxies for which a long term monitoring by MAGIC, VERITAS and HAWC was shown \citep{MoleroGonzalez:2023aez}. There were additional studies of the VHE $\gamma$-ray Propagation \citep{Cecil:2023bdq} and morphology \citep{BarbosaMartins:2023xcf}, for which the extension upper limits exclude the radio lobes as the origin of the VHE emission in the low state. Centaurus A is another of these sources for which the GeV gamma rays support a jet scenario, but hard X-rays are consistent with jet and corona models \citep{Rodi:2023vsv}. In general, for this type of sources, the comparison of the X-ray spectra of GeV-loud/quiet ones show that GeV-loud ones have a steeper spectrum and jets are less inclined \citep{Kayanoki:2023pxx}.

\subsection{Gamma Ray Bursts}

The merging of two compact objects or a hypernova are the origin of the so-called Gamma Ray Bursts (GRBs). In this conference, we had the luck of being less than one year away from GRB 221009A, the brightest of all time (B.O.A.T.). 

Several GRBs have been detected in the last few years at TeV energies while we were searching for them since many years ago. The possibility that this distribution has only a chance origin is discussed in \cite{Ashkar:2023ixb}. The reality is that at the moment, five GRBs have been detected at VHE gamma rays: GRB 180720B (H.E.S.S.), GRB 190114C (MAGIC), GRB 190829A (H.E.S.S.), GRB 201216C (MAGIC), GRB 221009A (LHAASO), all of them are long duration GRBs (duration T90 > 2 sec) and what it has been detected is the afterglow emission very likely produced by Synchroton Self Compton (SSC) emission by relativistic electrons in the forward shock.

\begin{figure}
   \centering
   \includegraphics[width=0.6\hsize]{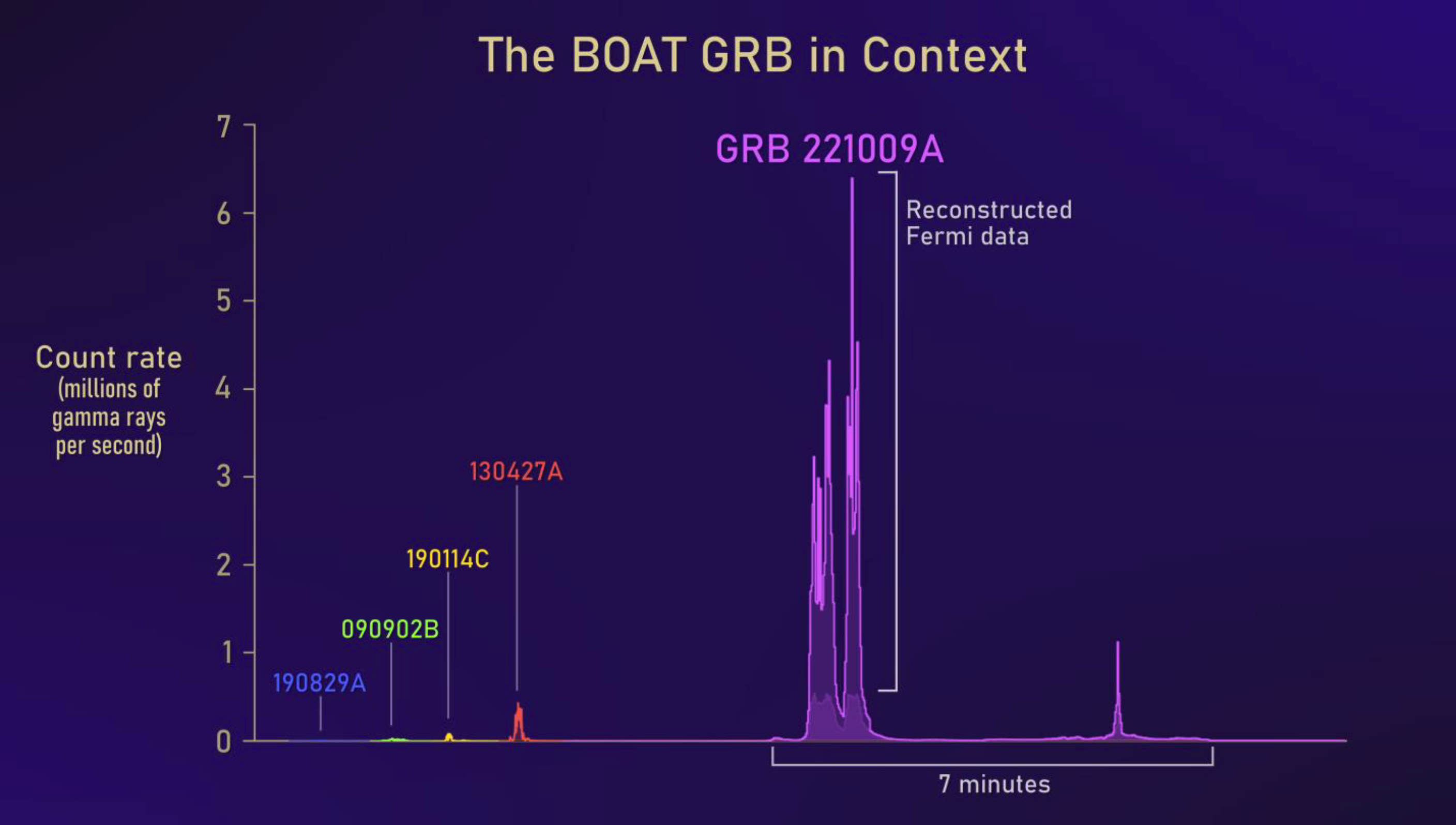}
   \includegraphics[width=0.39\hsize]{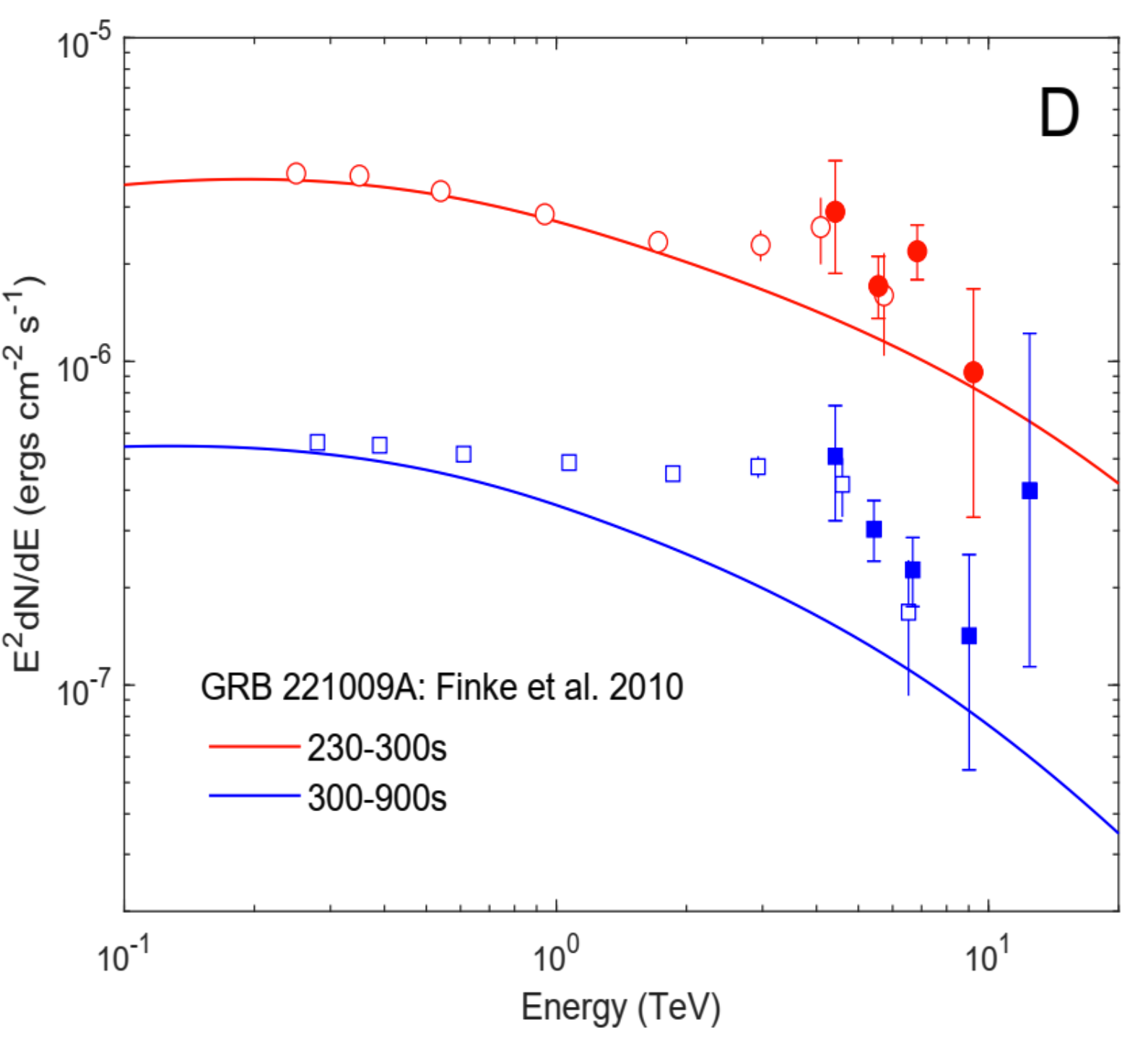}   
\caption{GRB 221009A count rate (left panel) and LHAASO spectra for different integration times after the burst (right panel).}
\label{fig:BOAT}
\end{figure}

The B.O.A.T. mentioned above (GRB 221009A) showed no polarized emission as measured by IXPE \cite{Dilalla:2023bsm}, breaking records (see Fig. \ref{fig:BOAT}) in {\it Fermi}-GBM \cite{Lesage:2023vme} and {\it Fermi}-LAT \citep{Bissaldi:2023yzz} being the one with the highest isotropic energy, the highest fluence, the highest peak flux and the 3rd highest isotropic intrinsic luminosity. No detection by IACTs \cite{HESS:2023yeg} and a detection by LHAASO that puts a lot of constraints \cite{Wang:2023ljn}. LHAASO detected photons up to 13 TeV from the afterglow \citep{Wang:2023ljn} putting the most stringent limits on the prompt TeV emission (emission detected only 230 s after the alert) that could mean that either there is no SSC emission or the absorption is too high. LHAASO data above 3 TeV also hints to an additional component as it is seen on the right panel of Fig. \ref{fig:BOAT}. Due to the energies reached by this emission, it is difficult to explain $\geq$ 10 TeV leptonic emission due to SSC \citep{Das:2023vhi}. Thanks to these observations, there were constraints put on the emission of Ultra High Energy CRs from GRBs \citep{Zhang:2023bjg}, limits derived on the Intergalactic Magnetic Field (IGMF) ($>10^{-18}$~G \citep{Huang:2023eii} or $4\times10^{-14}$ G \citep{Xia:2023mbe}, depending on the assumption) and even dark matter constraints \citep{Gonzalez:2023exp}.

We also had theory presentations like that of the study of the origin of afterglow plateaus with a promising interpretation coming from structured jet \citep{Stratta:2023nhr}. Models for the afterglow emission of GRBs based on GRB 190114C \citep{Asano:2023evh, Mondal:2023hhk}, constraints on the IGMF \cite{Vovk:2023wma} using GRB190114C data.

Finally, further GRB searches were presented like that from LHAASO-WCDA \citep{Huang:2023etf} for which only upper limits were derived so far except for GRB 221009A. There were also contributions about the monitoring of SDSS 1430+2303, the first candidate host for the merger of a Supermassive Black Hole Binary with H.E.S.S. \citep{HESS:2023otq} with no detection so far, the {\it Fermi}-LAT detection of the afterglow of GRB 211211A \citep{Zhang:2023evh} or the CTAO prospects comparing different CTAO Medium-size telescopes array layouts performances for gamma-ray burst observations \citep{CTASCTProject:2023ejz}.

\subsection{Cosmology}

The Extragalactic Background Light (EBL) is the light emitted by galaxies after the re-ionization era. This EBL has been measured using results from HE and VHE gamma-ray observations from distant AGNs and computing the $\gamma$-ray attenuation of their spectra. In this conference we had a proposal for a new EBL measurement, this time using a Bayesian mathematical framework \cite{Greaux:2023cay} that in general agrees with previous estimations except at low wavelengths. Using this EBL measurements, one can perform other cosmological estimations, such as for example to put upper limits on the redshift of galaxies \cite{Lainez:2023enu} or the measurement of the Hubble constant \cite{Kirkeberg:2023enu}. Additionally, using the spectra from several AGNs, the first EBL skymap was built to study its anisotropies \cite{Hervet:2023wtn} for which no significant dipole is observed at the moment, but shows promising results for the time of the advent of CTAO when more precise measurements will be performed.

\section{Surveys, Software and analysis methods}
There were several surveys and catalogs covering a wide energy range that were presented in this conference. We could see the presentation of several results obtained using satellites like INTEGRAL, COMPTEL and {\it Fermi}-LAT \cite{Orlando:2023jhs}, DAMPE with almost 250 sources in 7 years of data \cite{DAMPE:2023nxu} or CALET with an improved $\gamma$-ray reconstruction \cite{CALET:2023vld}. We also saw the preparation for the second H.E.S.S. Galactic Plane Survey catalogue \cite{Remy:2023xzj} and the second HAWC high energy catalog \cite{HAWC:2023amj} apart from the presentation of LHAASO catalogs below \cite{Hu:2023ljn} and above 25 TeV \cite{Xi:2023yng}.

The software development is also essential for the advancement of Science, in particular in fields in which the analysis of Big Data is a must. There were several presentations about methods, simulations and analysis software like ctapipe \cite{Linhoff:2023atg}, the low-level data processing pipeline software for CTAO, gammapy \cite{Khelifi:2023pky} the high level one, pyirf \cite{CTAConsortium:2023kub} the software to build the instrument response functions in CTAO, astripipe \cite{Mastropietro:2023qmb} the ASTRI pipeline, NectarChain \cite{CTANectarCAMProject:2023odx} Nectar one or cosipy \cite{Martinez-Castellanos:2023gse}, the COSI pipeline or libraries to fit data (Bjet\_MCMC) \cite{Hervet:2023wdb} or catalogs (STeVECat \cite{Greaux:2023xes}). It is also worth noting the role that deep learning is starting to acquire, not only to improve the data processing, but also in any task related to population classification or regression of some quantity, for which we had more than ten contributions in the conference.



\section{Final remarks}
From the results shown at the conference, my personal take is that we currently are in a transitional period: we have been having a couple of decades of great data that have been surprising us and telling us in which direction to move forward with the models, theory and phenomenology, but we need to take a leap step to be able to disprove all these models that are now at hand. There are a some exceptions like a few instruments coming into operation that are already producing great results, and also transient events that are always surprising us.
But we are lucky, the future is very bright, in the next few years there are projected experiments and observatories that will boost the covering of the $\gamma$-ray sky with an unprecedented sensitivity that will imply an expansion in the theories explaining all these new observations.

{\scriptsize	
\section*{Acknowledgements}
I would like to start thanking the SOC and LOC of this conference for giving me the opportunity of giving this talk and for the great organization of such an important and overwhelming conference, the level has been set very high for future organizers. I would also like to acknowledge the Ram\'on y Cajal program through grant RYC-2020-028639-I. He also acknowledges financial support from the Spanish "Ministerio de Ciencia e Innovación" (MCIN/AEI/ 10.13039/501100011033) through the Center of Excellence Severo Ochoa award for the Instituto de Astrofísica de Andalucía-CSIC (CEX2021-001131-S), and through grants PID2019-107847RB-C44 and PID2022-139117NB-C44.MCIN/AEI/10.13039/501100011033. 

\bibliographystyle{JHEP}
\bibliography{references}

\providecommand{\href}[2]{#2}\begingroup\raggedright\begin{thebibliography}{100}

\bibitem{Produit:2023rup}
N.~Produit, H.~Li and R.~Walter, \emph{{The Cosmic X-ray Background explorer (CXBe)}}, \href{https://doi.org/10.22323/1.444.0911}{\emph{PoS} {\bfseries ICRC2023} (2023) 911}.

\bibitem{Produit:2023dei}
N.~Produit et~al., \emph{{POLAR-2, the next generation of GRB polarization detector}}, \href{https://doi.org/10.22323/1.444.0550}{\emph{PoS} {\bfseries ICRC2023} (2023) 550} [\href{https://arxiv.org/abs/2309.00518}{{\ttfamily 2309.00518}}].

\bibitem{Fletcher:2023zxp}
C.~Fletcher, M.~Hui and A.~Goldstein, \emph{{The Scientific Performance of the The Moon Burst Energetics All-sky Monitor (MoonBEAM)}}, \href{https://doi.org/10.22323/1.444.0953}{\emph{PoS} {\bfseries ICRC2023} (2023) 953} [\href{https://arxiv.org/abs/2308.16293}{{\ttfamily 2308.16293}}].

\bibitem{Uchida:2023qhk}
Y.~Uchida et~al., \emph{{Performance of the XL-Calibur flight in July 2022 }}, \href{https://doi.org/10.22323/1.444.0928}{\emph{PoS} {\bfseries ICRC2023} (2023) 928}.

\bibitem{Tomsick:2023aue}
J.A.~Tomsick et~al., \emph{{The Compton Spectrometer and Imager}}, \href{https://doi.org/10.22323/1.444.0745}{\emph{PoS} {\bfseries ICRC2023} (2023) 745} [\href{https://arxiv.org/abs/2308.12362}{{\ttfamily 2308.12362}}].

\bibitem{Valverde:2023cyb}
J.~Valverde et~al., \emph{{The Compton-Pair telescope: A prototype for a next-generation MeV $\gamma$-ray observatory}}, \href{https://doi.org/10.22323/1.444.0857}{\emph{PoS} {\bfseries ICRC2023} (2023) 857} [\href{https://arxiv.org/abs/2308.12464}{{\ttfamily 2308.12464}}].

\bibitem{Karwin:2023xbg}
C.~Karwin et~al., \emph{{AMEGO-X: Concept and Future Plans }}, \href{https://doi.org/10.22323/1.444.0622}{\emph{PoS} {\bfseries ICRC2023} (2023) 622}.

\bibitem{Moiseev:2023zkv}
A.~Moiseev, \emph{{New Mission Concept: Compton Telescope with Coded Aperture Mask (GECCO) for MeV Gamma-ray Astronomy}}, \href{https://doi.org/10.22323/1.444.0702}{\emph{PoS} {\bfseries ICRC2023} (2023) 702}.

\bibitem{Ikeda:2023efo}
T.~Ikeda et~al., \emph{{Results of the SMILE-2+ balloon experiment}}, \href{https://doi.org/10.22323/1.444.0663}{\emph{PoS} {\bfseries ICRC2023} (2023) 663}.

\bibitem{Aramaki:2023ubm}
T.~Aramaki et~al., \emph{{The GRAMS (Gamma-Ray and AntiMatter Survey) Project}}, \href{https://doi.org/10.22323/1.444.0868}{\emph{PoS} {\bfseries ICRC2023} (2023) 868}.

\bibitem{Okuma:2023fnk}
K.~Okuma et~al., \emph{{Development of miniSGD, a proof-of-concept balloon experiment for a narrow field of view Si/CdTe semiconductor Compton telescope}}, \href{https://doi.org/10.22323/1.444.0899}{\emph{PoS} {\bfseries ICRC2023} (2023) 899}.

\bibitem{Lega:2023iuj}
A.~Lega, F.M.~Follega, R.~Iuppa, M.~Mese, R.~Nicolaidis, F.~Nozzoli et~al., \emph{{Exploring the Efficiency of HEPD-02 LYSO:Ce Scintillators in the CSES-02 Satellite Mission for Detecting Gamma-Ray Bursts}}, \href{https://doi.org/10.22323/1.444.0758}{\emph{PoS} {\bfseries ICRC2023} (2023) 758}.

\bibitem{Hashizume:2023ijl}
M.~Hashizume, \emph{{Performance Evaluation of \textquotedblleft{}XRPIX\textquotedblright{} Event-Driven SOI Pixel Detector for Cosmic MeV Gamma-ray Observation}}, \href{https://doi.org/10.22323/1.444.0873}{\emph{PoS} {\bfseries ICRC2023} (2023) 873}.

\bibitem{CALET:2023vld}
M.~Mori et~al., \emph{{Results from CALorimetric Electron Telescope (CALET) Observations of Gamma-rays on the International Space Station}}, \href{https://doi.org/10.22323/1.444.0708}{\emph{PoS} {\bfseries ICRC2023} (2023) 708}.

\bibitem{DAMPE:2023amf}
Z.-Q.~Shen et~al., \emph{{Recent progresses on the $\gamma$-ray observations of DAMPE}}, \href{https://doi.org/10.22323/1.444.0670}{\emph{PoS} {\bfseries ICRC2023} (2023) 670}.

\bibitem{Laviron:2023rhx}
A.~Laviron, D.~Bernard and P.~Bruel, \emph{{MeV-GeV polarimetry with the $Fermi$-LAT}}, \href{https://doi.org/10.22323/1.444.0721}{\emph{PoS} {\bfseries ICRC2023} (2023) 721}.

\bibitem{Takahashi:2023hgs}
S.~Takahashi, \emph{{GRAINE, balloon-borne emulsion telescope experiments for precise gamma-ray observations}}, \href{https://doi.org/10.22323/1.444.0598}{\emph{PoS} {\bfseries ICRC2023} (2023) 598}.

\bibitem{HERD:2023zjq}
G.~Lucchetta et~al., \emph{{Gamma-ray performance of the High Energy cosmic-Radiation Detection (HERD) space mission}}, \href{https://doi.org/10.22323/1.444.0691}{\emph{PoS} {\bfseries ICRC2023} (2023) 691}.

\bibitem{Zhang:2023zjq}
Y.~Zhang et~al., \emph{{Previous studies about BGO calorimeter in Very Large Area gamma ray Space Telescope}}, \href{https://doi.org/10.22323/1.444.0897}{\emph{PoS} {\bfseries ICRC2023} (2023) 897}.

\bibitem{Chen:2023nij}
W.~Chen and J.H.~Buckley, \emph{{Simulation of the instrument performance of the Antarctic Demonstrator for the Advanced Particle-astrophysics Telescope in the presence of the MeV background}}, \href{https://doi.org/10.22323/1.444.0841}{\emph{PoS} {\bfseries ICRC2023} (2023) 841}.

\bibitem{MAGIC:2023liw}
D.~Paneque et~al., \emph{{The MAGIC of VHE gamma-ray astronomy: 20 years, 200 peer-reviewed publications and beyond}}, \href{https://doi.org/10.22323/1.444.0624}{\emph{PoS} {\bfseries ICRC2023} (2023) 624}.

\bibitem{Giuliani:2023smc}
A.~Giuliani, G.~Pareschi, S.~Scuderi, S.~Lombardi and S.~Vercellone, \emph{{Status of the ASTRI program: technology and science with wide-field aplanatic IACT telescopes}}, \href{https://doi.org/10.22323/1.444.0892}{\emph{PoS} {\bfseries ICRC2023} (2023) 892}.

\bibitem{CTALSTProject:2023vhk}
D.~Mazin et~al., \emph{{First Science Results from CTA LST-1 Telescope and status of LST2-4}}, \href{https://doi.org/10.22323/1.444.0731}{\emph{PoS} {\bfseries ICRC2023} (2023) 731}.

\bibitem{CTALSTProject:2023ezj}
D.~Morcuende et~al., \emph{{Performance of the Large-Sized Telescope prototype of the Cherenkov Telescope Array}}, \href{https://doi.org/10.22323/1.444.0594}{\emph{PoS} {\bfseries ICRC2023} (2023) 594}.

\bibitem{Bradascio:2023lmr}
F.~Bradascio, \emph{{Status of the Medium-Sized Telescopes for the Cherenkov Telescope Array Observatory}}, \href{https://doi.org/10.22323/1.444.0859}{\emph{PoS} {\bfseries ICRC2023} (2023) 859}.

\bibitem{Kieda:2023upg}
D.~Kieda, \emph{{Status and Upgrades of the Schwarzschild-Couder Medium-Sized Telescope for the Cherenkov Telescope Array}}, \href{https://doi.org/10.22323/1.444.0642}{\emph{PoS} {\bfseries ICRC2023} (2023) 642}.

\bibitem{CTAConsortium:2023ngo}
A.~Trois et~al., \emph{{Status of the Small-Sized Telescopes Programme for the Cherenkov Telescope Array}}, \href{https://doi.org/10.22323/1.444.0917}{\emph{PoS} {\bfseries ICRC2023} (2023) 917}.

\bibitem{Antonelli:2023uig}
L.A.~Antonelli, \emph{{The CTA LST South as part of the CTA+ program}}, \href{https://doi.org/10.22323/1.444.0755}{\emph{PoS} {\bfseries ICRC2023} (2023) 755}.

\bibitem{Qian:2023xzx}
X.~Qian, H.~Sun and X.~Wang, \emph{{Prospective study on observations of $\gamma$-ray sources using the HADAR experiment}}, \href{https://doi.org/10.22323/1.444.0943}{\emph{PoS} {\bfseries ICRC2023} (2023) 943}.

\bibitem{DiPierro:2023olc}
F.~Di~Pierro, A.~Berti, R.~de~Menezes, E.~Jobst, Y.~Ohtani, J.~Sitarek et~al., \emph{{Performance of joint gamma-ray observations with MAGIC and LST-1 telescopes}}, \href{https://doi.org/10.22323/1.444.0636}{\emph{PoS} {\bfseries ICRC2023} (2023) 636}.

\bibitem{CTALSTProject:2023mpu}
J.~Baxter et~al., \emph{{Low energy performance boost through a hardware stereoscopic trigger between CTA LST1 and MAGIC}}, \href{https://doi.org/10.22323/1.444.0700}{\emph{PoS} {\bfseries ICRC2023} (2023) 700}.

\bibitem{CTAConsortium:2023ybn}
A.~Donini, I.~Burelli, O.~Gueta, F.~Longo, E.~Pueschel, D.~Tak et~al., \emph{{Performance study update of observations in divergent mode for the Cherenkov Telescope Array}}, \href{https://doi.org/10.22323/1.444.0840}{\emph{PoS} {\bfseries ICRC2023} (2023) 840}.

\bibitem{Zhang:2023bsi}
S.~Zhang, \emph{{Research and Development of Large Array of imaging atmospheric Cherenkov Telescopes}}, \href{https://doi.org/10.22323/1.444.0808}{\emph{PoS} {\bfseries ICRC2023} (2023) 808}.

\bibitem{Heller:2023qbh}
M.~Heller et~al., \emph{{The next generation cameras for the Large-Sized Telescopes of the Cherenkov Telescope Array Observatory}}, \href{https://doi.org/10.22323/1.444.0740}{\emph{PoS} {\bfseries ICRC2023} (2023) 740}.

\bibitem{Kieda:2023uig}
D.~Kieda, \emph{{Recent Observations by the VERITAS Stellar Intensity Interferometry (VSII) System}}, \href{https://doi.org/10.22323/1.444.0628}{\emph{PoS} {\bfseries ICRC2023} (2023) 628}.

\bibitem{MAGIC:2023ike}
I.~Jimenez-Martinez et~al., \emph{{Update on the performance of the MAGIC Intensity Interferometer}}, \href{https://doi.org/10.22323/1.444.0728}{\emph{PoS} {\bfseries ICRC2023} (2023) 728}.

\bibitem{HAWC:2023qbx}
M.~Mostafa et~al., \emph{{Recent Results From Hawc}}, \href{https://doi.org/10.22323/1.444.0756}{\emph{PoS} {\bfseries ICRC2023} (2023) 756}.

\bibitem{Gao:2023jgz}
B.~Gao et~al., \emph{{The running status of LHAASO-WCDA }}, \href{https://doi.org/10.22323/1.444.0800}{\emph{PoS} {\bfseries ICRC2023} (2023) 800}.

\bibitem{TibetASgamma:2023nks}
S.~Kato et~al., \emph{{Observation of gamma rays from the northern celestial sky up to the sub-PeV range with the Tibet air shower array and its underground muon detector array}}, \href{https://doi.org/10.22323/1.444.0627}{\emph{PoS} {\bfseries ICRC2023} (2023) 627}.

\bibitem{Pattanaik:2023mva}
D.~Pattanaik et~al., \emph{{Search for point sources of gamma-rays above 50 TeV with the GRAPES-3 experiment}}, \href{https://doi.org/10.22323/1.444.0922}{\emph{PoS} {\bfseries ICRC2023} (2023) 922}.

\bibitem{TAIGA:2023xtz}
T.~Ravdandorj et~al., \emph{{Gamma-hadron separation approach to point-like source observations in the TAIGA-IACT experiment in stereo observation mode}}, \href{https://doi.org/10.22323/1.444.0686}{\emph{PoS} {\bfseries ICRC2023} (2023) 686}.

\bibitem{SubietaVasquez:2023kci}
M.A.~Subieta~Vasquez, \emph{{Overview status of the ALPACA experiment}}, \href{https://doi.org/10.22323/1.444.0767}{\emph{PoS} {\bfseries ICRC2023} (2023) 767}.

\bibitem{Sako:2023kyj}
T.~Sako, \emph{{Mega ALPACA to explore multi-PeV gamma-ray sky in the southern hemisphere}}, \href{https://doi.org/10.22323/1.444.0632}{\emph{PoS} {\bfseries ICRC2023} (2023) 632}.

\bibitem{Conceicao:2023tfb}
R.~Concei\c{c}\~ao, \emph{{The Southern Wide-field Gamma-ray Observatory}}, \href{https://doi.org/10.22323/1.444.0963}{\emph{PoS} {\bfseries ICRC2023} (2023) 963} [\href{https://arxiv.org/abs/2309.04577}{{\ttfamily 2309.04577}}].

\bibitem{Korzoun:2023jgb}
N.~Korzoun et~al., \emph{{PeV Gamma-ray Astronomy With Panoramic Optical SETI Telescopes}}, \href{https://doi.org/10.22323/1.444.0787}{\emph{PoS} {\bfseries ICRC2023} (2023) 787} [\href{https://arxiv.org/abs/2308.09607}{{\ttfamily 2308.09607}}].

\bibitem{Fang:2023ugh}
K.~Fang, \emph{{TAIGA-IACT telescopes for the Multi-Messenger observations}}, \href{https://doi.org/10.22323/1.444.0560}{\emph{PoS} {\bfseries ICRC2023} (2023) 560}.

\bibitem{Chen:2023tpu}
S.~Chen, K.~Fang, Z.~Li and C.~LIU, \emph{{Detection of Gamma-ray emission from W51 region with LHAASO}}, \href{https://doi.org/10.22323/1.444.0957}{\emph{PoS} {\bfseries ICRC2023} (2023) 957}.

\bibitem{HAWC:2023sfg}
R.~Turner et~al., \emph{{The Boomerang PWN and its SNR G106.3+2.7 Viewed in the Very-High-Energy Gamma-Ray Regime by the HAWC Observatory}}, \href{https://doi.org/10.22323/1.444.0693}{\emph{PoS} {\bfseries ICRC2023} (2023) 693}.

\bibitem{MAGIC:2023kkn}
M.~Strzys et~al., \emph{{Highlights of Galactic Science with the MAGIC telescopes}}, \href{https://doi.org/10.22323/1.444.0955}{\emph{PoS} {\bfseries ICRC2023} (2023) 955}.

\bibitem{Park:2023bsm}
N.~Park, \emph{{VERITAS very high energy gamma-ray observations of SNR G106.3+2.7 region}}, \href{https://doi.org/10.22323/1.444.0869}{\emph{PoS} {\bfseries ICRC2023} (2023) 869}.

\bibitem{Feng:2023daw}
Y.~Feng et~al., \emph{{Study of Gamma-Ray Emission of the $\gamma$-Cygni SNR(G78.2+2.1) with LHAASO}}, \href{https://doi.org/10.22323/1.444.0602}{\emph{PoS} {\bfseries ICRC2023} (2023) 602}.

\bibitem{Hou:2023rte}
B.~Hou et~al., \emph{{Discovery of the Ultra-High-Energy gamma-ray source LHAASO J2002+3238 spatially associated with SNR G69.7+1.0 }}, \href{https://doi.org/10.22323/1.444.0605}{\emph{PoS} {\bfseries ICRC2023} (2023) 605}.

\bibitem{Zeng:2023uvu}
H.~Zeng, Y.~Guo, H.~Wu, Y.~Su, S.~Liu and Y.~Zhang, \emph{{High energy gamma-ray emission detected by LHAASO from SNR G150.3+4.5}}, \href{https://doi.org/10.22323/1.444.0606}{\emph{PoS} {\bfseries ICRC2023} (2023) 606}.

\bibitem{Oka:2023vaw}
T.~Oka, \emph{{Detection of delayed gamma-ray emissions around supernova remnant HB9 using Fermi- LAT observations}}, \href{https://doi.org/10.22323/1.444.0830}{\emph{PoS} {\bfseries ICRC2023} (2023) 830}.

\bibitem{Clement:2023sbx}
A.~Cl\'ement, M.~Lemoine-Goumard, V.~Wakelam, P.~Slane and P.~Gratier, \emph{{Constraints on cosmic-ray interaction and propagation within the HB3/W3 complex using Fermi-LAT data}}, \href{https://doi.org/10.22323/1.444.0578}{\emph{PoS} {\bfseries ICRC2023} (2023) 578}.

\bibitem{Giuffrida:2023udf}
R.~Giuffrida, M.~Lemoine-Goumard, M.~Miceli, S.~Gabici, Y.~Fukui and S.~Hidetoshi, \emph{{Evidence for proton acceleration and escape from the Puppis A SNR using Fermi-LAT observations}}, \href{https://doi.org/10.22323/1.444.0647}{\emph{PoS} {\bfseries ICRC2023} (2023) 647} [\href{https://arxiv.org/abs/2308.14848}{{\ttfamily 2308.14848}}].

\bibitem{Aruga:2023udf}
M.~Aruga et~al., \emph{{Molecular and atomic clouds associated with the gamma-ray supernova remnant Puppis A }}, \href{https://doi.org/10.22323/1.444.0935}{\emph{PoS} {\bfseries ICRC2023} (2023) 935}.

\bibitem{Einecke:2023njn}
S.~Einecke, G.P.~Rowell, J.~Pilossof, M.~Burton and K.O.~Cubuk, \emph{{Modelling the Gamma-ray Morphology of the Supernova Remnant W28}}, \href{https://doi.org/10.22323/1.444.0923}{\emph{PoS} {\bfseries ICRC2023} (2023) 923}.

\bibitem{Rowell:2023ypc}
G.~Rowell, S.~Einecke, R.~K\"onig, R.~Burley, P.~Marinos, M.~Filipovic et~al., \emph{{Modeling cosmic rays escaping from RXJ1713.7-3946 - Is it a really a PeVatron?}}, \href{https://doi.org/10.22323/1.444.0676}{\emph{PoS} {\bfseries ICRC2023} (2023) 676}.

\bibitem{Celli:2023nqg}
S.~Celli, A.~Specovius, A.~Mitchell, G.~Morlino and S.~Menchiari, \emph{{Gamma-ray emission from molecular clouds illuminated by local young massive stellar clusters and detection prospects with current and next generation instruments}}, \href{https://doi.org/10.22323/1.444.0775}{\emph{PoS} {\bfseries ICRC2023} (2023) 775}.

\bibitem{Harer:2023xdq}
L.~H\"arer, \emph{{Particle acceleration in superbubbles: MHD simulations and $\gamma$-ray signatures}}, \href{https://doi.org/10.22323/1.444.0854}{\emph{PoS} {\bfseries ICRC2023} (2023) 854}.

\bibitem{Mitchell:2023sxq}
A.M.W.~Mitchell, S.~Celli, A.~Specovius, G.~Morlino and S.~Menchiari, \emph{{Searching for evidence of PeVatron activity from stellar clusters via gamma-ray and neutrino signatures}}, \href{https://doi.org/10.22323/1.444.0611}{\emph{PoS} {\bfseries ICRC2023} (2023) 611}.

\bibitem{Guevel:2023bhg}
D.J.~Guevel et~al., \emph{{Limits on Leptonic TeV Emission from the Cygnus Cocoon with Swift-XRT}}, \href{https://doi.org/10.22323/1.444.0799}{\emph{PoS} {\bfseries ICRC2023} (2023) 799}.

\bibitem{Wu:2023ljn}
S.~Wu, \emph{{Highlight of LHAASO science results on PeVatrons}}, \href{https://doi.org/10.22323/1.444.0010}{\emph{PoS} {\bfseries ICRC2023} (2023) 010}.

\bibitem{Holch:2023jhb}
T.L.~Holch and E.~de~O\~na Wilhelmi, \emph{{An updated view of the VHE gamma-ray sky around the stellar cluster Westerlund 2 with the H.E.S.S. experiment}}, \href{https://doi.org/10.22323/1.444.0778}{\emph{PoS} {\bfseries ICRC2023} (2023) 778}.

\bibitem{Wang:2023dsw}
G.~Wang et~al., \emph{{Observation of the gamma-ray Emission from the W43 Direction with LHAASO}}, \href{https://doi.org/10.22323/1.444.0603}{\emph{PoS} {\bfseries ICRC2023} (2023) 603}.

\bibitem{2016Natur.531..476H}
{HESS Collaboration}, A.~{Abramowski}, F.~{Aharonian}, F.A.~{Benkhali}, A.G.~{Akhperjanian}, E.O.~{Ang{\"u}ner} et~al., \emph{{Acceleration of petaelectronvolt protons in the Galactic Centre}}, \href{https://doi.org/10.1038/nature17147}{\emph{Nature} {\bfseries 531} (2016) 476} [\href{https://arxiv.org/abs/1603.07730}{{\ttfamily 1603.07730}}].

\bibitem{CTALSTProject:2023njo}
H.~Abe et~al., \emph{{Joint Observation of the Galactic Center with MAGIC and CTA-LST-1}}, \href{https://doi.org/10.22323/1.444.0573}{\emph{PoS} {\bfseries ICRC2023} (2023) 573}.

\bibitem{Dorner:2023rul}
J.~D\"orner, J.~Becker~Tjus, P.-S.~Blomenkamp, H.~Fichtner, A.~Franckowiak, M.~Hoerbe et~al., \emph{{Implications from 3-dimensional modelling of gamma-ray signatures in the Galactic Center}}, \href{https://doi.org/10.22323/1.444.0584}{\emph{PoS} {\bfseries ICRC2023} (2023) 584}.

\bibitem{Ryan:2023yzu}
J.L.~Ryan, \emph{{Search for Dark Matter Annihilation Signals in the Galactic Center Halo with VERITAS}}, \href{https://doi.org/10.22323/1.444.0794}{\emph{PoS} {\bfseries ICRC2023} (2023) 794}.

\bibitem{DAMPE:2023aff}
Z.-Q.~Shen et~al., \emph{{Analysis of the Galactic center excess with DAMPE }}, \href{https://doi.org/10.22323/1.444.0671}{\emph{PoS} {\bfseries ICRC2023} (2023) 671}.

\bibitem{Zhang:2023xxp}
Y.~Zhang, \emph{{Investigation of the Ultra High Energy gamma-ray emission from the North Fermi Bubble with LHAASO-KM2A}}, \href{https://doi.org/10.22323/1.444.0651}{\emph{PoS} {\bfseries ICRC2023} (2023) 651}.

\bibitem{Li:2023dpg}
H.~Li, P.~Zhang, S.~Hu, M.~Zha, Y.~Guo, Q.~Yuan et~al., \emph{{Measurement of very-high-energy diffuse gamma-ray emission from |b|\ensuremath{<}5\textdegree{} degree of the Galactic plane with LHAASO-WCDA}}, \href{https://doi.org/10.22323/1.444.0672}{\emph{PoS} {\bfseries ICRC2023} (2023) 672}.

\bibitem{Zhang:2023caw}
R.~Zhang et~al., \emph{{Measurement of ultra-high-energy diffuse gamma-ray emission from |b|$<5^\circ$ of the Galactic plane with LHAASO-KM2A}}, \href{https://doi.org/10.22323/1.444.0610}{\emph{PoS} {\bfseries ICRC2023} (2023) 610}.

\bibitem{DelaTorreLuque:2023usg}
P.~De~la Torre~Luque, D.~Gaggero, D.~Grasso and A.~Marinelli, \emph{{Galactic diffuse gamma rays meet the PeV frontier}}, \href{https://doi.org/10.22323/1.444.0563}{\emph{PoS} {\bfseries ICRC2023} (2023) 563}.

\bibitem{Kaci:2023iie}
S.~Kaci and G.~Giacinti, \emph{{Galactic diffuse gamma-ray emission at VHE from discrete CR sources}}, \href{https://doi.org/10.22323/1.444.0773}{\emph{PoS} {\bfseries ICRC2023} (2023) 773}.

\bibitem{Giacinti:2023ljr}
G.~Giacinti, M.~Kachelriess, S.~Koldobskiy, A.~Neronov and D.~Semikoz, \emph{{Model for the diffuse gamma-ray and neutrino emission of the Milky Way at multi-TeV energies}}, \href{https://doi.org/10.22323/1.444.0813}{\emph{PoS} {\bfseries ICRC2023} (2023) 813}.

\bibitem{Stall:2023hns}
A.~Stall, L.~Kaiser and P.~Mertsch, \emph{{Stochastic modelling of cosmic ray sources for diffuse high-energy gamma-rays and neutrinos}}, \href{https://doi.org/10.22323/1.444.0687}{\emph{PoS} {\bfseries ICRC2023} (2023) 687} [\href{https://arxiv.org/abs/2309.02860}{{\ttfamily 2309.02860}}].

\bibitem{Abounnasr:2023ufg}
T.~Abounnasr, \emph{{Extended gamma-ray sources from anisotropic diffusion around PeVatrons}}, \href{https://doi.org/10.22323/1.444.0583}{\emph{PoS} {\bfseries ICRC2023} (2023) 583}.

\bibitem{Zhang:2023zww}
P.-P.~Zhang, X.-Y.~He, W.~Liu, Y.-Q.~Guo and Q.~Yuan, \emph{{Interpretation of new measurements of B/C and diffuse gamma rays using freshly accelerated cosmic rays interacting with surrounding medium}}, \href{https://doi.org/10.22323/1.444.0814}{\emph{PoS} {\bfseries ICRC2023} (2023) 814}.

\bibitem{Menchiari:2023qim}
S.~Menchiari, G.~Morlino, E.~Amato and N.~Bucciantini, \emph{{Diffuse $\gamma$-ray emission from a synthetic Galactic population of young stellar clusters}}, \href{https://doi.org/10.22323/1.444.0649}{\emph{PoS} {\bfseries ICRC2023} (2023) 649}.

\bibitem{Mizuno:2023qyf}
T.~Mizuno, H.~Ohno, E.~Watanabe, N.~Bucciantini, S.~Gunji, S.~Shibata et~al., \emph{{IXPE view of the Crab pulsar wind nebula}}, \href{https://doi.org/10.22323/1.444.0630}{\emph{PoS} {\bfseries ICRC2023} (2023) 630}.

\bibitem{Giacinti:2023tde}
G.~Giacinti, B.~Reville and J.~Kirk, \emph{{Origin of the PeV photons from the Crab Nebula}}, \href{https://doi.org/10.22323/1.444.0839}{\emph{PoS} {\bfseries ICRC2023} (2023) 839}.

\bibitem{Spencer:2023oqh}
S.~Spencer, A.~Mitchell and B.~Reville, \emph{{Hadronic re-acceleration at the Crab pulsar wind termination shock as a source of PeV gamma-rays}}, \href{https://doi.org/10.22323/1.444.0690}{\emph{PoS} {\bfseries ICRC2023} (2023) 690}.

\bibitem{Tsirou:2023fge}
M.~Tsirou et~al., \emph{{Studying the flaring emission of the Crab pulsar wind nebula system in high-energy gamma-rays }}, \href{https://doi.org/10.22323/1.444.0831}{\emph{PoS} {\bfseries ICRC2023} (2023) 831}.

\bibitem{You:2023jtj}
X.~You, S.~Hu and S.~Xi, \emph{{Updated analysis of the brightest UHE Gamma-ray Source LHAASO J1825-1326}}, \href{https://doi.org/10.22323/1.444.0643}{\emph{PoS} {\bfseries ICRC2023} (2023) 643}.

\bibitem{HAWC:2023fdd}
D.~Huang et~al., \emph{{Revealing Ultra-High-Energy Gamma-Ray Emission from the eHWC J1825-134 Region with HAWC}}, \href{https://doi.org/10.22323/1.444.0796}{\emph{PoS} {\bfseries ICRC2023} (2023) 796}.

\bibitem{Herzog:2023wmj}
I.~Herzog, \emph{{A Spectral, Morphological, and Emission Analysis of Gamma Ray Source HAWC J2031+415}}, \href{https://doi.org/10.22323/1.444.0757}{\emph{PoS} {\bfseries ICRC2023} (2023) 757}.

\bibitem{Li:2023ryl}
B.~Li, \emph{{Modelling the X-Ray Emission from the Magnetar Wind Nebula around Swift J1834.9-0846}}, \href{https://doi.org/10.22323/1.444.0895}{\emph{PoS} {\bfseries ICRC2023} (2023) 895}.

\bibitem{Bi:2023nxr}
X.-J.~Bi, \emph{{Slow diffusion around pulsar $\gamma$-ray halos and its impact on cosmic rays propagation}},  in \emph{{38th International Cosmic Ray Conference}}, 8, 2023 [\href{https://arxiv.org/abs/2308.08099}{{\ttfamily 2308.08099}}].

\bibitem{Liu:2023cjn}
R.-Y.~Liu, Q.-Z.~Wu and K.~Yan, \emph{{Particle Transport in Pulsar Halos and Their Contribution to the Diffuse Galactic Gamma-ray Emission}}, \href{https://doi.org/10.22323/1.444.0833}{\emph{PoS} {\bfseries ICRC2023} (2023) 833}.

\bibitem{Wu:2023lgn}
Q.~Wu et~al., \emph{{Diagnosing the particle transport mechanism in the Geminga,Monogem and PSR J0622+3749's pulsar halo via X-ray observation }}, \href{https://doi.org/10.22323/1.444.0735}{\emph{PoS} {\bfseries ICRC2023} (2023) 735}.

\bibitem{2017Sci...358..911A}
A.U.~{Abeysekara}, A.~{Albert}, R.~{Alfaro}, C.~{Alvarez}, J.D.~{{\'A}lvarez}, R.~{Arceo} et~al., \emph{{Extended gamma-ray sources around pulsars constrain the origin of the positron flux at Earth}}, \href{https://doi.org/10.1126/science.aan4880}{\emph{Science} {\bfseries 358} (2017) 911} [\href{https://arxiv.org/abs/1711.06223}{{\ttfamily 1711.06223}}].

\bibitem{HAWC:2023bfh}
R.~Torres~Escobedo et~al., \emph{{TeV Halo Study of Geminga and Monogem with HAWC}}, \href{https://doi.org/10.22323/1.444.0710}{\emph{PoS} {\bfseries ICRC2023} (2023) 710}.

\bibitem{Mitchell:2023rrd}
A.M.W.~Mitchell and S.~Caroff, \emph{{Modelling of highly extended Gamma-ray emission around the Geminga Pulsar as detected with H.E.S.S}}, {\emph{PoS} {\bfseries ICRC2023} (2023) 590} [\href{https://arxiv.org/abs/2308.16669}{{\ttfamily 2308.16669}}].

\bibitem{Chen:2023ffo}
E.~Chen et~al., \emph{{Study of particle diffusion around Geminga with LHAASO-KM2A}},  in \emph{{38th International Cosmic Ray Conference}}, vol.~ICRC2023, p.~613, 2023, \href{https://doi.org/10.22323/1.444.0613}{DOI}.

\bibitem{HESS:2023owq}
T.~Wach et~al., \emph{{Joint H.E.S.S. and Fermi-LAT analysis of the region around PSR J1813-1749}}, \href{https://doi.org/10.22323/1.444.0589}{\emph{PoS} {\bfseries ICRC2023} (2023) 589} [\href{https://arxiv.org/abs/2308.16717}{{\ttfamily 2308.16717}}].

\bibitem{Xu:2023jtl}
R.F.~Xu and K.~Wang, \emph{{Possible VHE gamma emission from PWN tail of PSR J1740+1000}}, \href{https://doi.org/10.22323/1.444.0607}{\emph{PoS} {\bfseries ICRC2023} (2023) 607}.

\bibitem{Coutino:2023fsf}
S.~Couti\~no~de Leon, \emph{{HAWC Observation of TeV halos around gamma-ray pulsars }}, \href{https://doi.org/10.22323/1.444.0793}{\emph{PoS} {\bfseries ICRC2023} (2023) 793}.

\bibitem{Coutino:2023gsf}
S.~Couti\~no~de Leon, \emph{{HAWC Detection of a TeV Halo Candidate Surrounding PSR J0359+5414}}, \href{https://doi.org/10.22323/1.444.0792}{\emph{PoS} {\bfseries ICRC2023} (2023) 792}.

\bibitem{HAWC:2023bfj}
Q.~Wu et~al., \emph{{A Stacking Search for Tev Halos Around Millisecond Pulsars With HAWC}}, \href{https://doi.org/10.22323/1.444.0735}{\emph{PoS} {\bfseries ICRC2023} (2023) 735}.

\bibitem{HAWC:2023fsz}
B.~Andres et~al., \emph{{Constraining the TeV halo population in M31}}, \href{https://doi.org/10.22323/1.444.0768}{\emph{PoS} {\bfseries ICRC2023} (2023) 768}.

\bibitem{Eckner:2023jiz}
C.~Eckner, \emph{{Detecting and characterizing pulsar halos with the Cherenkov Telescope Array}}, \href{https://doi.org/10.22323/1.444.0772}{\emph{PoS} {\bfseries ICRC2023} (2023) 772}.

\bibitem{Li:2023fkj}
B.~Li et~al., \emph{{Prospect of detecting X-ray haloes around middle-aged pulsars with eROSITA}}, \href{https://doi.org/10.22323/1.444.0570}{\emph{PoS} {\bfseries ICRC2023} (2023) 570}.

\bibitem{CTALSTProject:2023ekj}
K.~Abe et~al., \emph{{First results of pulsar observations with the LST-1}}, \href{https://doi.org/10.22323/1.444.0569}{\emph{PoS} {\bfseries ICRC2023} (2023) 569}.

\bibitem{Li:2023uvu}
Z.~Li et~al., \emph{{ The Ultra-high-energy emission towards millisecond pulsar J0218+4232 by LHAASO}}, \href{https://doi.org/10.22323/1.444.0609}{\emph{PoS} {\bfseries ICRC2023} (2023) 609}.

\bibitem{Razzano:2023gnk}
M.~Razzano et~al., \emph{{Multiwavelength observations of the variable gamma-ray pulsar PSR J2021+4026 }}, \href{https://doi.org/10.22323/1.444.0898}{\emph{PoS} {\bfseries ICRC2023} (2023) 898}.

\bibitem{Olivera:2023ljn}
L.~Olivera~Nieto, \emph{{Transport of relativistic particles observed in the parsec-scale jets of SS 433}}, \href{https://doi.org/10.22323/1.444.0011}{\emph{PoS} {\bfseries ICRC2023} (2023) 011}.

\bibitem{HAWC:2023dtw}
C.~Rho et~al., \emph{{Spectral Study of the West Jet Lobe of SS 433 with HAWC}}, \href{https://doi.org/10.22323/1.444.0769}{\emph{PoS} {\bfseries ICRC2023} (2023) 769}.

\bibitem{Thorpe:2023lrg}
C.~Thorpe-Morgan et~al., \emph{{H.E.S.S. Observations of the 2021 PSR B1259-63 Periastron }}, \href{https://doi.org/10.22323/1.444.0585}{\emph{PoS} {\bfseries ICRC2023} (2023) 585}.

\bibitem{Fisher:2023bdg}
L.~Fisher et~al., \emph{{H.E.S.S. Observations of the Gamma-ray Binary LMC P3 }}, \href{https://doi.org/10.22323/1.444.0906}{\emph{PoS} {\bfseries ICRC2023} (2023) 906}.

\bibitem{HESS:2023nna}
S.~Steinmassl et~al., \emph{{The Eta Carinae 2020 periastron passage as seen by H.E.S.S.}}, \href{https://doi.org/10.22323/1.444.0640}{\emph{PoS} {\bfseries ICRC2023} (2023) 640}.

\bibitem{Walter:2023hfl}
R.~Walter and M.~Balbo, \emph{{Modelling the $\gamma$-ray emission from $\eta$ Carinae}}, \href{https://doi.org/10.22323/1.444.0909}{\emph{PoS} {\bfseries ICRC2023} (2023) 909}.

\bibitem{HAWC:2023acx}
X.~Wang et~al., \emph{{HAWC observations of microquasars as powerful particle accelerators}}, \href{https://doi.org/10.22323/1.444.0797}{\emph{PoS} {\bfseries ICRC2023} (2023) 797}.

\bibitem{Holder:2023tlg}
J.~Holder, \emph{{VERITAS observations of the Be/X-ray binary system LS V +44 17 during a major outburst.}}, \href{https://doi.org/10.22323/1.444.0695}{\emph{PoS} {\bfseries ICRC2023} (2023) 695} [\href{https://arxiv.org/abs/2308.12214}{{\ttfamily 2308.12214}}].

\bibitem{MAGIC:2023wxo}
D.~Green et~al., \emph{{Evidence of hadronic origin of the gamma-ray emission from the nova RS Oph by the MAGIC telescopes}}, \href{https://doi.org/10.22323/1.444.0580}{\emph{PoS} {\bfseries ICRC2023} (2023) 580}.

\bibitem{CTALSTProject:2023mgl}
Y.~Kobayashi et~al., \emph{{Detection of the 2021 Outburst of RS Ophiuchi with the LST-1}}, \href{https://doi.org/10.22323/1.444.0677}{\emph{PoS} {\bfseries ICRC2023} (2023) 677}.

\bibitem{Diesing:2023chh}
R.~Diesing, B.D.~Metzger, E.~Aydi, L.~Chomiuk, I.~Vurm, S.~Gupta et~al., \emph{{Using Gamma-Rays to Reveal the Evolution of Novae}}, \href{https://doi.org/10.22323/1.444.0865}{\emph{PoS} {\bfseries ICRC2023} (2023) 865}.

\bibitem{Hadasch:2023ljn}
D.~Hadasch, \emph{{Exploring the Universe's Extreme Events: Galactic Transients at High and Very High Energies}}, \href{https://doi.org/10.22323/1.444.0022}{\emph{PoS} {\bfseries ICRC2023} (2023) 022}.

\bibitem{HAWC:2023gcx}
M.~Lundy et~al., \emph{{Latest Fast Radio Burst Observations with VERITAS }}, \href{https://doi.org/10.22323/1.444.0798}{\emph{PoS} {\bfseries ICRC2023} (2023) 798}.

\bibitem{Jaitly:2023fcx}
A.~Jaitly et~al., \emph{{Searching for Short Timescale Transients in Gamma-ray Telescope Data}}, \href{https://doi.org/10.22323/1.444.0861}{\emph{PoS} {\bfseries ICRC2023} (2023) 861}.

\bibitem{Willox:2023lsf}
E.~Willox, \emph{{Search for TeV Gamma Ray Emission from Fast Radio Burst Locations with the HAWC Obsevatory}}, \href{https://doi.org/10.22323/1.444.0596}{\emph{PoS} {\bfseries ICRC2023} (2023) 596}.

\bibitem{Ashkar:2023omc}
H.~Ashkar, M.E.~Bouhaddouti, S.~Fegan and F.~Sch\"ussler, \emph{{All sky archival search for FRB high energy counterparts with Swift and Fermi}}, \href{https://doi.org/10.22323/1.444.0555}{\emph{PoS} {\bfseries ICRC2023} (2023) 555} [\href{https://arxiv.org/abs/2309.02883}{{\ttfamily 2309.02883}}].

\bibitem{Kleiner:2023ehi}
T.K.~Kleiner, \emph{{VERITAS Observations of MGRO J1908+06}}, \href{https://doi.org/10.22323/1.444.0762}{\emph{PoS} {\bfseries ICRC2023} (2023) 762}.

\bibitem{Bangale:2023ktt}
P.~Bangale and X.~Wang, \emph{{Searching for TeV emission from LHAASO J0341+5258 with VERITAS and HAWC}}, \href{https://doi.org/10.22323/1.444.0706}{\emph{PoS} {\bfseries ICRC2023} (2023) 706} [\href{https://arxiv.org/abs/2308.15643}{{\ttfamily 2308.15643}}].

\bibitem{Yu:2023jia}
Y.~Yu et~al., \emph{{Observation the UHE Gamma-ray emission from the J1959+2850 with LHAASO }}, \href{https://doi.org/10.22323/1.444.0621}{\emph{PoS} {\bfseries ICRC2023} (2023) 621}.

\bibitem{Guo:2023jia}
Y.~Guo et~al., \emph{{Probing the origin of UHE $\gamma$-ray emission from LHAASO J1929+1745 region with updated LHAASO data}}, \href{https://doi.org/10.22323/1.444.0608}{\emph{PoS} {\bfseries ICRC2023} (2023) 608}.

\bibitem{CTALSTProject:2023bfh}
G.~Pirola et~al., \emph{{Multi-wavelength analysis of the PeVatron candidate LHAASO J2108+5157}}, \href{https://doi.org/10.22323/1.444.0727}{\emph{PoS} {\bfseries ICRC2023} (2023) 727}.

\bibitem{Kumar:2023txz}
S.~Kumar, M.~Martin and X.~Wang, \emph{{VERITAS and HAWC observations of unidentified source LHAASO J2108+5157}}, \href{https://doi.org/10.22323/1.444.0941}{\emph{PoS} {\bfseries ICRC2023} (2023) 941} [\href{https://arxiv.org/abs/2309.00089}{{\ttfamily 2309.00089}}].

\bibitem{Toledano:2023ncz}
I.~Toledano\textendash{}Ju\'arez et~al., \emph{{Unveiling the molecular environment of the enigmatic PeVatron candidate LHAASO J2108+5157}}, \href{https://doi.org/10.22323/1.444.0809}{\emph{PoS} {\bfseries ICRC2023} (2023) 809}.

\bibitem{HAWC:2023dsl}
Y.~Son et~al., \emph{{Study of the HAWC counterpart of LHAASO J1849-0003 and its surroundings at TeV energies}}, \href{https://doi.org/10.22323/1.444.0581}{\emph{PoS} {\bfseries ICRC2023} (2023) 581}.

\bibitem{HAWC:2023lst}
R.~Babu et~al., \emph{{High energy gamma ray emission from HESSJ1809-193: Morphological and Spectral studies with HAWC}}, \href{https://doi.org/10.22323/1.444.0789}{\emph{PoS} {\bfseries ICRC2023} (2023) 789}.

\bibitem{Li:2023mia}
C.-M.~Li, C.~Ge and R.-Y.~Liu, \emph{{X-ray observation of HESS J1809-193: indication of an X-ray halo and implication for its gamma-ray origin}}, \href{https://doi.org/10.22323/1.444.0561}{\emph{PoS} {\bfseries ICRC2023} (2023) 561}.

\bibitem{Saha:2023ent}
L.~Saha, \emph{{VERITAS Observations of M 82 and Other Selected Starburst Galaxies}}, \href{https://doi.org/10.22323/1.444.0746}{\emph{PoS} {\bfseries ICRC2023} (2023) 746}.

\bibitem{Owen:2023sqe}
E.R.~Owen, A.K.H.~Kong and K.-C.~Pan, \emph{{Cosmic ray calorimetry in star-forming galaxy populations and implications for their contribution to the extra-galactic $\gamma$-ray background}}, \href{https://doi.org/10.22323/1.444.0554}{\emph{PoS} {\bfseries ICRC2023} (2023) 554} [\href{https://arxiv.org/abs/2308.04793}{{\ttfamily 2308.04793}}].

\bibitem{Baghmanyan:2023aqf}
V.~Baghmanyan, L.~Oswald, L.~Pfeiffer, A.~Azzollini, E.~Barbano, G.F.~de et~al., \emph{{Investigating the nature of TeV gamma-ray variability in blazars}}, \href{https://doi.org/10.22323/1.444.0921}{\emph{PoS} {\bfseries ICRC2023} (2023) 921}.

\bibitem{Lindfors:2023efw}
E.~Lindfors et~al., \emph{{Fast variability of the VHE gamma-rays; quantitative comparison with magnetic reconnection models}}, \href{https://doi.org/10.22323/1.444.0905}{\emph{PoS} {\bfseries ICRC2023} (2023) 905}.

\bibitem{Boula:2023upv}
S.~Boula, \emph{{Investigating the Physical Mechanisms of Blazar Emission: A Multi-Zone Emission Model}}, \href{https://doi.org/10.22323/1.444.0890}{\emph{PoS} {\bfseries ICRC2023} (2023) 890}.

\bibitem{Dmytriiev:2023wkf}
A.~Dmytriiev and M.~Boettcher, \emph{{Effects of non-continuous inverse Compton cooling in blazars}}, \href{https://doi.org/10.22323/1.444.0645}{\emph{PoS} {\bfseries ICRC2023} (2023) 645}.

\bibitem{Boughelilba:2023fet}
M.~Boughelilba, A.~Reimer, L.~Merten and J.P.~Lundquist, \emph{{Spine-sheath jet model for low-luminosity AGNs}}, \href{https://doi.org/10.22323/1.444.0958}{\emph{PoS} {\bfseries ICRC2023} (2023) 958} [\href{https://arxiv.org/abs/2308.10596}{{\ttfamily 2308.10596}}].

\bibitem{Tomar:2023wzn}
G.~Tomar, N.~Gupta and R.~Prince, \emph{{Emission from the jets of Low-luminosity Active Galactic Nuclei}}, \href{https://doi.org/10.22323/1.444.0947}{\emph{PoS} {\bfseries ICRC2023} (2023) 947}.

\bibitem{Heckmann:2023qtg}
L.~Heckmann, D.~Paneque and A.~Reimer, \emph{{A novel approach to identify blazar emission states using clustering algorithms}}, \href{https://doi.org/10.22323/1.444.0634}{\emph{PoS} {\bfseries ICRC2023} (2023) 634}.

\bibitem{CTAConsortium:2023grr}
M.~Cerruti et~al., \emph{{Bright blazar flares with CTA}}, \href{https://doi.org/10.22323/1.444.0850}{\emph{PoS} {\bfseries ICRC2023} (2023) 850}.

\bibitem{Grolleron:2023dtw}
G.~Grolleron et~al., \emph{{Variability studies of active galactic nuclei from the long-term monitoring program with the Cherenkov Telescope Array}}, \href{https://doi.org/10.22323/1.444.0856}{\emph{PoS} {\bfseries ICRC2023} (2023) 856}.

\bibitem{Lindfors:2023ffw}
E.~Lindfors et~al., \emph{{Recent results from the redshift determination of blazars for the Cherenkov Telescope Array}}, \href{https://doi.org/10.22323/1.444.0910}{\emph{PoS} {\bfseries ICRC2023} (2023) 910}.

\bibitem{Imazawa:2023ezy}
R.~Imazawa, J.~Stri\v{s}kovi\'c, J.~Jormanainen, S.~Truzzi, E.~Lindfors, D.~Dominis~Prester et~al., \emph{{MAGIC observation of BL Lacertae flaring period in 2020}}, \href{https://doi.org/10.22323/1.444.0595}{\emph{PoS} {\bfseries ICRC2023} (2023) 595}.

\bibitem{CTALSTProject:2023gmi}
K.~Abe et~al., \emph{{LST-1 observations of an enormous flare of BL Lacertae in 2021}}, \href{https://doi.org/10.22323/1.444.0552}{\emph{PoS} {\bfseries ICRC2023} (2023) 552}.

\bibitem{Hinrichs:2023vjm}
C.E.~Hinrichs, A.~Acharyya and A.C.~Sadun, \emph{{Multi-wavelength Observations of a Long-duration Flare from BL Lacertae}}, \href{https://doi.org/10.22323/1.444.0747}{\emph{PoS} {\bfseries ICRC2023} (2023) 747}.

\bibitem{CTALSTProject:2023aoh}
K.~Abe et~al., \emph{{Observation of Active Galactic Nuclei with the Large-Sized Telescope prototype of the Cherenkov Telescope Array}}, \href{https://doi.org/10.22323/1.444.0711}{\emph{PoS} {\bfseries ICRC2023} (2023) 711}.

\bibitem{HAWC:2023szn}
J.A.~Garcia~Gonzalez et~al., \emph{{Comparing HAWC blazars light curves with different data reconstruction versions}}, \href{https://doi.org/10.22323/1.444.0889}{\emph{PoS} {\bfseries ICRC2023} (2023) 889}.

\bibitem{Mooney:2023ppi}
C.L.~Mooney, \emph{{A Precise and Long-term Correlation Study of X-ray and TeV Emission from Mrk 421}}, \href{https://doi.org/10.22323/1.444.0888}{\emph{PoS} {\bfseries ICRC2023} (2023) 888}.

\bibitem{Damico:2023rup}
G.~D'Amico et~al., \emph{{Constraints on Lorentz Invariance Violation using the extraordinary flare of Mrk 421 in 2014}}, \href{https://doi.org/10.22323/1.444.0912}{\emph{PoS} {\bfseries ICRC2023} (2023) 912}.

\bibitem{Becerra:2023lrb}
J.~Becerra~Gonzalez et~al., \emph{{Hunting for TeV Structured Jets}}, \href{https://doi.org/10.22323/1.444.0586}{\emph{PoS} {\bfseries ICRC2023} (2023) 586}.

\bibitem{Dominguez:2023enu}
A.~Dominguez et~al., \emph{{PG 1553+113: the case for a super-massive black hole binary system }}, \href{https://doi.org/10.22323/1.444.0559}{\emph{PoS} {\bfseries ICRC2023} (2023) 559}.

\bibitem{Benbow:2023maj}
W.~Benbow, \emph{{Recent Highlights from the VERITAS AGN Discovery Program}}, \href{https://doi.org/10.22323/1.444.0760}{\emph{PoS} {\bfseries ICRC2023} (2023) 760}.

\bibitem{Ribeiro:2023jgj}
D.~Ribeiro, \emph{{TeV Detection of the Extreme HSP Blazar RBS 1366 by VERITAS}}, \href{https://doi.org/10.22323/1.444.0659}{\emph{PoS} {\bfseries ICRC2023} (2023) 659}.

\bibitem{Acharyya:2023ogc}
A.~Acharyya, \emph{{The VERITAS discovery and multiwavelength observations of the blazar S3 1227+25}}, \href{https://doi.org/10.22323/1.444.0625}{\emph{PoS} {\bfseries ICRC2023} (2023) 625}.

\bibitem{HESS:2023mnr}
M.~Cerruti et~al., \emph{{Target of Opportunity Observations of Flaring Blazars with H.E.S.S.}}, \href{https://doi.org/10.22323/1.444.0924}{\emph{PoS} {\bfseries ICRC2023} (2023) 924} [\href{https://arxiv.org/abs/2308.07872}{{\ttfamily 2308.07872}}].

\bibitem{HAWC:2023gto}
F.J.~Urena~Mena et~al., \emph{{An updated survey of Active Galaxies with the HAWC gamma-ray observatory}}, \href{https://doi.org/10.22323/1.444.0805}{\emph{PoS} {\bfseries ICRC2023} (2023) 805}.

\bibitem{MAGIC:2023zrc}
H.A.~Mondal et~al., \emph{{Constraints on VHE gamma-ray emission of Flat Spectrum Radio Quasars with the MAGIC telescopes}}, \href{https://doi.org/10.22323/1.444.0777}{\emph{PoS} {\bfseries ICRC2023} (2023) 777}.

\bibitem{MAGIC:2023hrz}
S.~Loporchio et~al., \emph{{Multiwavelength characterization of two flaring blazars: insight into the emission region of intermediate-synchrotron-peaked BL Lacs}}, \href{https://doi.org/10.22323/1.444.0725}{\emph{PoS} {\bfseries ICRC2023} (2023) 725}.

\bibitem{MoleroGonzalez:2023aez}
M.~Molero~Gonzalez, L.~Fortson, M.~Nievas, E.~Pueschel, D.~Ribeiro, M.~V\'azquez et~al., \emph{{Long-term monitoring of the radio-galaxy M87 in gamma-rays: joint analysis of MAGIC, VERITAS and Fermi-LAT data}}, \href{https://doi.org/10.22323/1.444.0572}{\emph{PoS} {\bfseries ICRC2023} (2023) 572}.

\bibitem{Cecil:2023bdq}
R.~Cecil et~al., \emph{{Probing Gamma-Ray Propagation at Very-High Energies with H.E.S.S. Observations of M87 }}, \href{https://doi.org/10.22323/1.444.0908}{\emph{PoS} {\bfseries ICRC2023} (2023) 908}.

\bibitem{BarbosaMartins:2023xcf}
V.~Barbosa~Martins, S.~Ohm, C.~Arcaro, N.~\.Zywucka and d.N.~Mathieu, \emph{{Probing the morphology of the low state gamma-ray emission of M87 with H.E.S.S.}}, \href{https://doi.org/10.22323/1.444.0696}{\emph{PoS} {\bfseries ICRC2023} (2023) 696}.

\bibitem{Rodi:2023vsv}
J.~Rodi, E.~Jourdain, M.~Molina and J.-P.~Roques, \emph{{Investigating the Nature of the Hard X-ray/Soft Gamma-ray Emission from Centaurus A}}, \href{https://doi.org/10.22323/1.444.0650}{\emph{PoS} {\bfseries ICRC2023} (2023) 650}.

\bibitem{Kayanoki:2023pxx}
T.~Kayanoki and Y.~Fukazawa, \emph{{Relationship between Gamma-ray loudness and X-ray spectra of Radio Galaxies}}, \href{https://doi.org/10.22323/1.444.0567}{\emph{PoS} {\bfseries ICRC2023} (2023) 567}.

\bibitem{Ashkar:2023ixb}
H.~Ashkar, S.~Fegan and A.~Sangar\'e, \emph{{Why no VHE GRBs were detected before 2018?}}, \href{https://doi.org/10.22323/1.444.0638}{\emph{PoS} {\bfseries ICRC2023} (2023) 638}.

\bibitem{Dilalla:2023bsm}
N.~Di~Lalla, \emph{{The IXPE view of GRB 221009A }}, \href{https://doi.org/10.22323/1.444.0870}{\emph{PoS} {\bfseries ICRC2023} (2023) 870}.

\bibitem{Lesage:2023vme}
S.~Lesage et~al., \emph{{Fermi GBM Analysis of GRB 221009A }}, \href{https://doi.org/10.22323/1.444.0882}{\emph{PoS} {\bfseries ICRC2023} (2023) 882}.

\bibitem{Bissaldi:2023yzz}
E.~Bissaldi, P.~Bruel, N.~Omodei, R.~Pillera and N.~Di~Lalla, \emph{{GRB 221009A: The brightest burst of all time as seen by Fermi-LAT}}, \href{https://doi.org/10.22323/1.444.0847}{\emph{PoS} {\bfseries ICRC2023} (2023) 847}.

\bibitem{HESS:2023yeg}
J.D.~Mbarubucyeye et~al., \emph{{H.E.S.S. follow-up observations of GRB 221009A}}, \href{https://doi.org/10.22323/1.444.0705}{\emph{PoS} {\bfseries ICRC2023} (2023) 705}.

\bibitem{Wang:2023ljn}
X.~Wang, \emph{{LHAASO observations of the brightest-of-all-time GRB 221009A }}, \href{https://doi.org/10.22323/1.444.0009}{\emph{PoS} {\bfseries ICRC2023} (2023) 009}.

\bibitem{Das:2023vhi}
S.~Das and S.~Razzaque, \emph{{Cosmic-ray origin of $\gtrsim 10$ TeV gamma-rays in GRB\textasciitilde{}221009A}}, \href{https://doi.org/10.22323/1.444.0668}{\emph{PoS} {\bfseries ICRC2023} (2023) 668}.

\bibitem{Zhang:2023bjg}
X.-F.~Zhang, R.-Y.~Liu, H.-M.~Zhang, Y.-Y.~Huang, B.T.~Zhang and X.-Y.~Wang, \emph{{Constraints on Cosmic Rays acceleration of Bright Gamma-ray Bursts with Observations of Fermi-LAT}}, \href{https://doi.org/10.22323/1.444.0823}{\emph{PoS} {\bfseries ICRC2023} (2023) 823}.

\bibitem{Huang:2023eii}
Y.-Y.~Huang, H.-M.~Zhang, K.~Yan, R.-Y.~Liu and X.-Y.~Wang, \emph{{Detection of GeV emission from an ultralong gamma-ray burst with the Fermi Large Area Telescope}}, \href{https://doi.org/10.22323/1.444.0828}{\emph{PoS} {\bfseries ICRC2023} (2023) 828}.

\bibitem{Xia:2023mbe}
Z.~Xia et~al., \emph{{The inter-galactic magnetic field strength inferred with GRB 221009A }}, \href{https://doi.org/10.22323/1.444.0886}{\emph{PoS} {\bfseries ICRC2023} (2023) 886}.

\bibitem{Gonzalez:2023exp}
M.M.~Gonz\'alez, D.A.~Rojas, A.~Pratts, S.~Hern\'andez-Cadena, N.~Fraija, R.~Alfaro et~al., \emph{{GRB 221009A: A light dark matter burst or an extremely bright Inverse Compton component?}}, \href{https://doi.org/10.22323/1.444.0844}{\emph{PoS} {\bfseries ICRC2023} (2023) 844}.

\bibitem{Stratta:2023nhr}
G.~Stratta, \emph{{On the origin of afterglow “plateaus” in gamma-ray bursts: a broad-band spectro-temporal analysis }}, \href{https://doi.org/10.22323/1.444.0950}{\emph{PoS} {\bfseries ICRC2023} (2023) 950}.

\bibitem{Asano:2023evh}
K.~Asano, \emph{{Early Gamma-Ray Afterglow from Gamma-Ray Bursts}}, \href{https://doi.org/10.22323/1.444.0633}{\emph{PoS} {\bfseries ICRC2023} (2023) 633}.

\bibitem{Mondal:2023hhk}
T.~Mondal, L.~Resmi and D.~Bose, \emph{{Modeling of Gamma Ray Burst (GRB) Afterglow at Very High Energy (VHE) regime}}, \href{https://doi.org/10.22323/1.444.0600}{\emph{PoS} {\bfseries ICRC2023} (2023) 600}.

\bibitem{Vovk:2023wma}
I.~Vovk, \emph{{Search for pair echo signatures in the gamma-ray light curve of GRB190114C}}, \href{https://doi.org/10.22323/1.444.0712}{\emph{PoS} {\bfseries ICRC2023} (2023) 712}.

\bibitem{Huang:2023etf}
Y.~Huang, C.~Liu and Z.G.~Yao, \emph{{Search for very high energy gamma-ray bursts based on LHAASO-WCDA triggered data}}, \href{https://doi.org/10.22323/1.444.0826}{\emph{PoS} {\bfseries ICRC2023} (2023) 826}.

\bibitem{HESS:2023otq}
A.~Jaitly et~al., \emph{{Monitoring the first candidate host for the merger of a Supermassive Black Hole Binary with H.E.S.S.}}, \href{https://doi.org/10.22323/1.444.0688}{\emph{PoS} {\bfseries ICRC2023} (2023) 688}.

\bibitem{Zhang:2023evh}
H.~Zhang et~al., \emph{{Fermi-LAT Detection of a GeV Afterglow from a Compact Stellar Merger }}, \href{https://doi.org/10.22323/1.444.0887}{\emph{PoS} {\bfseries ICRC2023} (2023) 887}.

\bibitem{CTASCTProject:2023ejz}
D.~Cerasole et~al., \emph{{Comparative study of CTAO Medium-size telescopes array layouts performances in gamma-ray burst observations}}, \href{https://doi.org/10.22323/1.444.0726}{\emph{PoS} {\bfseries ICRC2023} (2023) 726}.

\bibitem{Greaux:2023cay}
L.~Gr\'eaux, J.~Biteau and M.~Nievas~Rosillo, \emph{{A gamma-ray perspective on the cosmological optical controversy}}, \href{https://doi.org/10.22323/1.444.0749}{\emph{PoS} {\bfseries ICRC2023} (2023) 749}.

\bibitem{Lainez:2023enu}
M.~L\'ainez, A.~Dom\'\i{}nguez, V.S.~Paliya, N.~\'Alvarez-Crespo, M.~Ajello, J.~Finke et~al., \emph{{Redshift estimates from extragalactic background light attenuation}}, \href{https://doi.org/10.22323/1.444.0558}{\emph{PoS} {\bfseries ICRC2023} (2023) 558}.

\bibitem{Kirkeberg:2023enu}
P.~Kirkeberg et~al., \emph{{A new derivation of the Hubble constant from gamma-ray attenuation using optical depths for the Fermi and CTA era }}, \href{https://doi.org/10.22323/1.444.0916}{\emph{PoS} {\bfseries ICRC2023} (2023) 916}.

\bibitem{Hervet:2023wtn}
O.~Hervet, D.~Williams and A.~Furniss, \emph{{The first skymap of the extragalactic background light from gamma-ray spectra, a new window on cosmological anisotropies}}, \href{https://doi.org/10.22323/1.444.0753}{\emph{PoS} {\bfseries ICRC2023} (2023) 753}.

\bibitem{Orlando:2023jhs}
E.~Orlando, \emph{{The Inner Galaxy at MeV with INTEGRAL and COMPTEL: Interstellar Emission or Sources?}}, \href{https://doi.org/10.22323/1.444.0896}{\emph{PoS} {\bfseries ICRC2023} (2023) 896}.

\bibitem{DAMPE:2023nxu}
K.-K.~Duan et~al., \emph{{Point-like Source Catalog Observed by DAMPE}}, \href{https://doi.org/10.22323/1.444.0669}{\emph{PoS} {\bfseries ICRC2023} (2023) 669}.

\bibitem{Remy:2023xzj}
Q.~Remy, \emph{{Towards the second H.E.S.S. Galactic Plane Survey catalogue}}, \href{https://doi.org/10.22323/1.444.0744}{\emph{PoS} {\bfseries ICRC2023} (2023) 744} [\href{https://arxiv.org/abs/2308.08969}{{\ttfamily 2308.08969}}].

\bibitem{HAWC:2023amj}
P.~Harding et~al., \emph{{The HAWC ultra-high-energy gamma-ray map with more than 5 years of data}}, \href{https://doi.org/10.22323/1.444.0698}{\emph{PoS} {\bfseries ICRC2023} (2023) 698}.

\bibitem{Hu:2023ljn}
S.C.~Hu, G.M.~Xiang, M.~Zha and Z.G.~Yao, \emph{{The First LHAASO Catalog of Gamma-ray Sources below 25TeV}}, \href{https://doi.org/10.22323/1.444.0655}{\emph{PoS} {\bfseries ICRC2023} (2023) 655}.

\bibitem{Xi:2023yng}
S.~Xi and S.~Chen, \emph{{LHAASO first \ensuremath{>}25 TeV Gamma-ray Source Catalog}}, \href{https://doi.org/10.22323/1.444.0620}{\emph{PoS} {\bfseries ICRC2023} (2023) 620}.

\bibitem{Linhoff:2023atg}
M.~Linhoff, L.~Beiske, N.~Biederbeck, S.~Fr\"ose, K.~Kosack and L.~Nickel, \emph{{ctapipe - Prototype Open Event Reconstruction Pipeline for the Cherenkov Telescope Array}}, \href{https://doi.org/10.22323/1.444.0703}{\emph{PoS} {\bfseries ICRC2023} (2023) 703}.

\bibitem{Khelifi:2023pky}
B.~Kh\'elifi, R.~Terrier, A.~Donath, A.~Sinha, Q.~Remy and F.~Pintore, \emph{{Gammapy: present status and future roadmap}}, \href{https://doi.org/10.22323/1.444.0959}{\emph{PoS} {\bfseries ICRC2023} (2023) 959} [\href{https://arxiv.org/abs/2308.13389}{{\ttfamily 2308.13389}}].

\bibitem{CTAConsortium:2023kub}
R.M.~Dominik et~al., \emph{{Interpolation of Instrument Response Functions for the Cherenkov Telescope Array in the Context of pyirf}}, \href{https://doi.org/10.22323/1.444.0618}{\emph{PoS} {\bfseries ICRC2023} (2023) 618}.

\bibitem{Mastropietro:2023qmb}
M.~Mastropietro, S.~Lombardi, F.~Lucarelli, F.~Visconti, E.~Fedorova and L.A.~Antonelli, \emph{{The ASTRI Mini-Array Cherenkov Data Pipeline}}, \href{https://doi.org/10.22323/1.444.0774}{\emph{PoS} {\bfseries ICRC2023} (2023) 774}.

\bibitem{CTANectarCAMProject:2023odx}
G.~Grolleron et~al., \emph{{NectarChain, the scientific software for the CTA-NectarCAM}}, \href{https://doi.org/10.22323/1.444.0862}{\emph{PoS} {\bfseries ICRC2023} (2023) 862}.

\bibitem{Martinez-Castellanos:2023gse}
I.~Martinez-Castellanos et~al., \emph{{The cosipy library: COSI\textquoteright{}s high-level analysis software}}, \href{https://doi.org/10.22323/1.444.0858}{\emph{PoS} {\bfseries ICRC2023} (2023) 858} [\href{https://arxiv.org/abs/2308.11436}{{\ttfamily 2308.11436}}].

\bibitem{Hervet:2023wdb}
O.~Hervet, C.~Johnson and A.~Youngquist, \emph{{Bjet\_MCMC: a new tool to automatically fit the broadband SEDs of blazars}}, \href{https://doi.org/10.22323/1.444.0754}{\emph{PoS} {\bfseries ICRC2023} (2023) 754}.

\bibitem{Greaux:2023xes}
L.~Gr\'eaux, J.~Biteau, T.~Hassan, O.~Hervet, M.~Nievas~Rosillo and D.A.~Williams, \emph{{STeVECat, the Spectral TeV Extragalactic Catalog}}, \href{https://doi.org/10.22323/1.444.0751}{\emph{PoS} {\bfseries ICRC2023} (2023) 751}.

\end{thebibliography}\endgroup
}

%
%
%

\end{document}